\documentclass[a4paper,11pt]{article}
\usepackage{jheppub} % for details on the use of the package, please see the JINST-author-manual
\usepackage{slashed}
\usepackage{multirow}
\usepackage[table]{xcolor}
\usepackage{float}

\usepackage{color}
\definecolor{darkgreen}{rgb}{0,0.5,0}

\usepackage{color}
\definecolor{royalblue}{rgb}{0.25,0.41,1}

\title{%Generalized symmetries and anomaly inflow \\
Generalized symmetries and emergence in axion effective field theories}

\author[a,b]{Nathaniel Craig}
\author[a]{and Dan Sehayek}
\affiliation[a]{Department of Physics, University of California, Santa Barbara, CA 93106, USA}
\affiliation[b]{Kavli Institute for Theoretical Physics, Santa Barbara, CA 93106, USA}

\emailAdd{dsehayek@ucsb.edu}

\abstract{We study the phenomenological consequences of higher symmetry structures in axion effective field theories. Higher-group and non-invertible symmetries impose parametric constraints on the energy scales at which different symmetries can emerge in the infrared, providing a guide to the ultraviolet physics. We clarify and analyze these emergence constraints in axion EFTs coupled to abelian and non-abelian gauge bosons, with and without charged matter. We show that emergence constraints are universally saturated by anomaly inflow onto topological defects, while in perturbative UV completions they are supererogatory owing to the parametric separation of scales.}

\begin{document}
\maketitle
\flushbottom

\section{Introduction}

The advent of \textit{generalized} or \textit{categorical} symmetries \cite{Baez2004,Baez2011,Gukov2013,Gaiotto2015,Kapustin2017,Cordova2019,Tachikawa2020,Cordova2021,emergent-hg-brennan,I9,SHS-non-invertible,Bhardwaj2023,Cordova-non-invertible,Bartsch2023,Bartsch2023-II,D3,SHS-non-invertible-II,Bhardwaj2023-II,Bhardwaj2023-III,Bhardwaj2024-II,I10,Bartsch2024-II,Bhardwaj2025} (see \cite{Cordova2022,brennan-hong,bhardwaj2023lecturesgeneralizedsymmetries,McGreevy,LUO2024,Schafernameki2023} for reviews) has had a profound impact across mathematics, high energy physics, and condensed matter physics, enabling (among other things) the successful generalization of the Landau paradigm. Within particle physics, they show promise in generalizing the 't Hooft paradigm\footnote{To paraphrase \cite{tHooft:1979rat} in analogy with the Landau paradigm, ``small parameters are classified by the symmetries they break.''} \cite{Tong2017,Gaiotto2018,Hidaka2020,Tanizaki2020,Hidaka2021,emergent-hg-brennan,SHS-non-invertible,Yokokura,Koren2022,Cordova-non-invertible,SHS-non-invertible-II,ReeceTwist,Choi2024,Cordova2024-II,cheung2024,Cordova2024-III,craig2024highqualityaxionshigherformsymmetries,Putrov2024,Koren2025,Cordova2025,Davighi2025}, although their phenomenological consequences have yet to be fully explored.

The archetypal example is the generalization of ordinary global symmetries to higher-form global symmetries \cite{Gaiotto2015,Kapustin2017,Seiberg-line-operators} acting on $p$-dimensional charged objects. A continuous $p$-form global symmetry is generally associated with a conserved $(p+1)$-index current $J^{\mu_1\cdots\mu_{p+1}}$ whose conservation implies the existence of a topological symmetry defect operator \footnote{Here, $J^{(p+1)}:=J_{\mu_1...\mu_{p+1}}dx^{\mu_1}\wedge\cdots\wedge dx^{\mu_{p+1}}$ and hence $d\star J^{(p+1)}=0$}, 
\begin{equation}
    U_\alpha(\Sigma_{d-p-1})=\exp\left [ i\alpha\int_{{\Sigma}_{d-p-1}}\star J^{(p+1)} \right ],
\end{equation}
which is invariant under smooth deformations of $\Sigma_{d-p-1}$. The collection of these symmetry defect operators and their relations can then be used to define the symmetry.

Perhaps the greatest phenomenological consequences for particle physics arise not from exact higher-form symmetries, but from anomalous ones. If a given global symmetry has an ABJ anomaly, it is still often possible to construct gauge-invariant topological symmetry operators involving the associated anomalous current by decorating these operators with appropriate topological quantum field theories (TQFTs). Since such defects are generally non-invertible, ABJ anomalous global symmetries are referred to as \textit{non-invertible} symmetries. Examples of non-invertible defects for ABJ anomalous symmetries can be found in \cite{SHS-non-invertible,SHS-non-invertible-II,Cordova-non-invertible,Yokokura,Aguilera,Damia2023,DelZotto:2024ngj}. Several more examples of non-invertible symmetries can be found in 1+1d conformal field theories (CFTs) and topological phases \cite{Verlinde1988,Petkova2001,Fuchs2002,Frohlich2004,Frohlich2007,Aasen2016,Bhardwaj2018,Chang2019,Lin2021,Thorngren2024,aasen2020topologicaldefectslatticedualities,Sharpe2023,Huang2021,Chang2023,Lin2023,seifnashri2024clusterstatenoninvertiblesymmetry}, including the non-invertible Kramers-Wannier duality of the Ising CFT \cite{KramersWannier,Shankar,shao2024whatsundonetasilectures}. Other recent examples of non-invertible symmetries in higher-dimensional theories, including lattice gauge theories and supersymmetric theories, have been discussed in \cite{Roumpedakis2023,Cordova2024,choi2024noninvertiblehigherformsymmetries21d,Argurio2023,I1,I2,I3,I4,I5,I6,I7,I8,I9,I10,Heckman2023,I11,I12,I13,I14}.

In addition to ABJ anomalies, a global symmetry in a given theory may have 't Hooft anomalies, which represent an obstruction to gauging the global symmetry. In some cases, a collection of symmetries that participate in a 't Hooft anomaly can still be consistently gauged by imposing a higher-group structure. Concretely, given a collection of $p_i$-form global symmetries with 
\begin{equation}
    p_1<p_2<\cdots<p_n,
\end{equation}
we can introduce a set of background gauge fields $B_i^{(p_i+1)}$ that would ordinarily transform as $\delta B_i^{(p_i+1)}=d\Lambda_i^{(p_i)}$ where $\Lambda_i^{(p_i)}$ is a $p_i$-form gauge parameter. We say that there is a $(p_n+1)$-group global symmetry if the theory is not invariant under the standard gauge transformations described above, but is instead invariant under a correlated set of background transformations of the form
\begin{equation}
    \delta B_i^{(p_i+1)}=d \Lambda_i^{(p_i)}+\sum_{j \leq i} \Lambda_j^{(p_j)} \wedge \alpha_j^{(i)}\left(\left\{B_j^{(p_j+1)}\right\}\right)+\mathcal{O}\left(\Lambda^2\right)
\end{equation}
where $\alpha_j^{(i)}$ is a $p_i-p_j+1$-form that depends on the background gauge fields $B_j^{(p_j+1)}$ for $j<i$ and the terms in $\mathcal{O}(\Lambda^2)$ are referred to as Schwinger terms. Well-known examples of such symmetries are the 2-group symmetries that arise from the gauging of magnetic 1-form symmetries \cite{Cordova2019} and the 3-group symmetries that arise from the mixed 't Hooft anomaly between the axion-periodicity and electric/center 1-form symmetry in axion-Maxwell/Yang-Mills \cite{emergent-hg-brennan,SHS-non-invertible-II,anomalies-coupling-constants,Hidaka2020,Hidaka2021}. See \cite{Kapustin2017,Benini2019,Barkeshli2018,Fidkowski2017,Hsin2020,Tanizaki2020,Iqbal2023,Cordova2021,Apruzzi2022,Hsin2022,DeWolfe2021,Delmastro2023,Barkeshli,BarkeshliII,Bhardwaj2024,Kang2024,Davighi:2024zjp} for additional examples of higher-group symmetries in topological phases, QCD, spin liquids, hydrodynamics, holography and superconformal field theories.

Both ABJ and 't Hooft anomalies of 0-form global symmetries have long provided a powerful guide to the phenomenology of gauge theories as a function of scale, since the anomalies must be matched along RG flows. Unsurprisingly, the higher symmetry structures induced by anomalies are equally powerful. In particular, both higher-group and non-invertible symmetries reveal constraints on the scales at which their constituent symmetries emerge. Generically, if a theory possesses an emergent $(p+1)$-form higher group symmetry that correlates background gauge field transformations of a non-anomalous $p$-form global symmetry with those of a 't Hooft-anomalous $q$-form global symmetry, then it follows that the emergence scale of the $p$-form symmetry must be parametrically greater than or equal to the emergence scale of the $q$-form symmetry. Furthermore, symmetry defects associated with non-invertible symmetries are generally constructed by half-gauging or half-higher-gauging a collection of symmetries, from which it follows that the emergence scale of the gauged symmetries must be parametrically greater than or equal to that of the non-invertible symmetry. 

%Phenomenology of emergence phenomena
Emergence constraints are a potentially powerful guide for phenomenology, as they must hold in any UV completion of an effective field theory with a higher symmetry structure. Such structures are ubiquitous in the Standard Model and its extensions, and some implications of emergence constraints have been explored in the context of Standard Model flavor symmetries \cite{Koren2022} and axion effective field theories \cite{emergent-hg-brennan,SHS-non-invertible-II}. The goal of this paper is to fully explore the physical phenomena associated with the satisfaction of emergence constraints in axion effective field theories, building on the results of  \cite{emergent-hg-brennan,SHS-non-invertible-II}. As we will see, the coupled background gauge redundancies implied by higher-form, higher-group, and non-invertible symmetries lead directly to parametric inequalities among the energy scales at which different symmetries can emerge. We show that these constraints are already visible at the level of the infrared effective description, independent of microscopic details, and that they admit a clear physical interpretation in terms of screening, dynamical charged objects, and the breakdown of topological symmetry operators. In perturbative ultraviolet completions, we find that the constraints are typically satisfied by a large parametric margin due to the scale separation implied by perturbativity. More generally, we demonstrate that anomaly inflow onto axion strings, monopoles, and their higher-dimensional analogues provides a universal, UV-independent mechanism that enforces all of the emergence constraints, tying together higher symmetry structure and infrared phenomenology.

%Outline
This paper is organized as follows: In Section~\ref{sec:am} we begin with axion--Maxwell theory as a minimal setting in which higher-form, higher-group, and non-invertible symmetries coexist, and we review the associated emergence constraints. We then interpret these constraints in a variety of perturbative UV completions, including four-dimensional KSVZ-type models and five-dimensional gauge theories, and sharpen their physical meaning using symmetry-breaking scales defined via effective charges. Section~\ref{sec:am} concludes with a general discussion of anomaly inflow mechanisms, which we show provide a universal explanation for the satisfaction of the emergence constraints. In Section~\ref{sec:aym} we extend the analysis to axion Yang--Mills theory, including different global structures of the gauge group, and derive the corresponding constraints. Section~\ref{sec:matter} discusses generalizations involving charged matter. We conclude in Section~\ref{sec:conc} with a summary and a discussion of open questions. Several appendices review background material on topological order, discrete and non-invertible symmetries in axion theories, and the emergence constraints implied by non-invertible defects.

\section{Axion-Maxwell} \label{sec:am}

\subsection{Generalized Symmetries and Emergence Constraints}

We begin with the conceptually simplest axion EFT, the axion-Maxwell theory of an axion coupled to a pure $U(1)$ gauge theory. (Here `Maxwell' indicates the absence of charged matter under the $U(1)$; we'll turn to axion-QED with charged matter in Sec.~\ref{sec:matter}.) Despite its conceptual simplicity, axion-Maxwell theory exhibits a rich set of symmetries, including higher-form, higher-group \cite{emergent-hg-brennan,Hidaka2021}, and non-invertible symmetries \cite{SHS-non-invertible,SHS-non-invertible-II,Cordova-non-invertible}. In Euclidean signature, it is described by the following action
\begin{eqnarray}
    S=\int_{\mathcal{M}_4}\frac{1}{2}da\wedge\star da+\frac{1}{2e_4^2}F\wedge\star F-\frac{iK}{8\pi^2f}a F\wedge F
\label{eq:axion-Maxwell-3+1d}
\end{eqnarray}
where $F=dA$ is the abelian field strength of a $U(1)$ gauge field $A$ and $a$ is the axion. Without the axion-gauge boson coupling, the theory enjoys both a continuous zero-form axion-shift symmetry $U(1)_s^{(0)}$ and a two-form winding symmetry $U(1)_w^{(2)}$ with respective conserved currents
\begin{align}
\star J^{(1)}=i f \star d a, \quad \star J^{(3)}=\frac{1}{2 \pi f} d a \, ,
\end{align}
as well as an electric 1-form symmetry $U(1)_e^{(1)}$ and magnetic 1-form symmetry $U(1)_m^{(1)}$ with respective conserved currents
\begin{align}
    \star J_{e}^{(2)}=\frac{i}{e_4^2}\star F,\quad \star J_{m}^{(2)}=\frac{F}{2 \pi} \, .
\end{align} 
In the presence of the axion-gauge boson coupling, the 0-form axion-shift and electric 1-form symmetries are broken down to discrete non-invertible symmetries \cite{SHS-non-invertible,SHS-non-invertible-II,Cordova-non-invertible} (see also Appendix \ref{sec:non-invertible}). This follows from the equations of motion,
\begin{align}
    d \star J^{(1)}=\frac{K}{8 \pi^2} F \wedge F, \quad d \star J_{e}^{(2)}=-\frac{K}{4 \pi^2 f} d(a F) \ . 
\end{align}
To diagnose the global symmetry structure of the axion–Maxwell theory, it is useful to couple the theory to background gauge fields for its invertible global symmetries -- in this case, the invertible $\mathbb{Z}_K$ subgroups of these non-invertible axion-shift and electric 1-form symmetries, as well as the continuous magnetic 1-form and 2-form winding symmetries. The $\mathbb{Z}_K$ symmetries are described by the symmetry operators $\mathcal{U}_\alpha(\Sigma_3)=\exp\left ( i\alpha Q(\Sigma_3)\right )$ and $\mathcal{U}_\alpha(\Sigma_2)=\exp\left ( i\alpha Q(\Sigma_2)\right )$ where
\begin{align}
    Q(\Sigma_3)=\int_{\Sigma_3}\star J^{(1)}-\frac{K}{8\pi^2}A\wedge F, \quad Q(\Sigma_2)=\int_{\Sigma_2}\star J_e^{(2)}+\frac{K}{4\pi^2f}aF \label{eq:charges}
\end{align}
and $\alpha=2\pi n/K$ for $n\in\mathbb{Z}$ is required for invariance under large gauge transformations of $A$ and axion-shifts $a\rightarrow a+2\pi f$, respectively. Moreover, the presence of the axion coupling to gauge bosons suggests a formulation in terms of anomaly inflow, which naturally leads to a five-dimensional bulk extension for this term. Upon gauging the invertible symmetries and extending the topological term to a five-dimensional manifold, Equation \ref{eq:axion-Maxwell-3+1d} becomes
\begin{align} \nonumber
S & =\frac{1}{2} \int_{\mathcal{M}_4}\left(d a-f X^{(1)}\right) \wedge \star\left(d a-f X^{(1)}\right)+\frac{1}{2 e_4^2} \int_{\mathcal{M}_4}\left(F-B_e^{(2)}\right) \wedge \star\left(F-B_e^{(2)}\right) \\
& +\frac{i}{2 \pi f} \int_{\mathcal{M}_4} a dC^{(3)}-\frac{i}{2 \pi} \int_{\mathcal{M}_4} dB_m^{(2)} \wedge A \\ \nonumber
&-\frac{iK}{8 \pi^2 f} \int_{\mathcal{M}_5}\left(d a-f X^{(1)}\right) \wedge\left(F-B_e^{(2)}\right) \wedge\left(F-B_e^{(2)}\right) \ .
\end{align}
where $\partial\mathcal{M}_5=\mathcal{M}_4$, $X^{(1)}$ and $B_e^{(2)}$ are $\mathbb{Z}_K$ background gauge fields for the axion-shift and electric 1-form symmetries, and $B_m^{(2)}$ and $C^{(3)}$ are background gauge fields for the continuous magnetic 1-form and 2-form winding symmetries. Furthermore, we have defined the topological term on a five-dimensional manifold $\mathcal{M}_5$ to ensure that the $2\pi$-periodicity of the axion is preserved. Namely, without such an extension, a $2\pi$-shift of the axion field generates a non-trivial phase given by $-\frac{iK}{4\pi}\int_{\mathcal{M}_4}B_e^{(2)}\wedge B_e^{(2)}\in \frac{2\pi i\mathbb{Z}}{K}$, which we refer to as a mixed 't Hooft anomaly between the $2\pi$-periodicity of the axion and the electric 1-form symmetry. 

The higher-group structure is revealed by noting that the physical theory on $\mathcal{M}_4$ cannot depend on the bulk extension. Namely, we require that
\begin{equation}
    \int_{\mathcal{M}_5 \cup_{\mathcal{M}_4} \overline{\mathcal{M}}_5^{\prime}}\left[\left(dC^{(3)}-\frac{K}{4 \pi} B_e^{(2)} \wedge B_e^{(2)}\right) \wedge d a+\left(dB_m^{(2)}-\frac{K}{2 \pi} X^{(1)} \wedge B_e^{(2)}\right) \wedge d A\right]=0\bmod2 \pi i \mathbb{Z}
\end{equation}
where $\cup_{\mathcal{M}_4}$ denotes gluing along the boundary of $\mathcal{M}_4$. Hence, the field strengths for the magnetic 1-form and 2-form winding symmetries must be modified as
\begin{align}
dC^{(3)}&\mapsto G^{(4)}=d C^{(3)}+\frac{K}{4 \pi} B_e^{(2)} \wedge B_e^{(2)} \label{eq:axion-Maxwell-G} \\
dB_m^{(2)}&\mapsto H^{(3)}=d B_m^{(2)}+\frac{K}{2 \pi} X^{(1)} \wedge B_e^{(2)} \label{eq:axion-Maxwell-H}
\end{align}
Invariance of these modified field strengths then requires the modified set of background gauge field transformations
\begin{align}
& C^{(3)} \longmapsto C^{(3)}+d \Lambda^{(2)}-\frac{K}{2 \pi} B_e^{(2)} \wedge d \Lambda_e^{(1)}-\frac{K}{4 \pi} \Lambda_e^{(1)} \wedge d \Lambda_e^{(1)} \label{eq:hg-maxwell-3+1d-a}\\
& B_m^{(2)} \longmapsto B_m^{(2)}+d \Lambda_m^{(1)}+\frac{K}{2 \pi} X^{(1)} \wedge \Lambda_e^{(1)}-\frac{K}{2 \pi} B_e^{(2)} \Lambda^{(0)}+\frac{K}{2 \pi} d \Lambda^{(0)} \wedge \Lambda_e^{(1)} \label{eq:hg-maxwell-3+1d-b}\\
& B_e^{(2)} \longmapsto B_e^{(2)}+d \Lambda_e^{(1)} \label{eq:hg-maxwell-3+1d-c} \\
& X^{(1)} \longmapsto X^{(1)} + d \Lambda^{(0)}
\end{align}
With this 3-group symmetry, the gauged action is fully invariant under all background gauge field transformations provided that the following inflow action is also present:
\begin{equation}
    S_{\text {inflow }}=\frac{i}{2 \pi} \int_{\mathcal{M}_5} G^{(4)} \wedge X^{(1)}+\frac{i}{2 \pi} \int_{\mathcal{M}_5} H^{(3)} \wedge B_e^{(2)} .
\end{equation}

With the modified background gauge transformations Equations \ref{eq:hg-maxwell-3+1d-a}--\ref{eq:hg-maxwell-3+1d-c} in hand, we can already see that the symmetry data are not factorized: the background gauge field for one symmetry does not transform independently of the others. This intertwining of background gauge redundancies is the defining signature of a higher-group structure, reflecting the fact that turning on a background for one symmetry necessarily requires correlated backgrounds and gauge transformations for the others.

The emergence conjecture translates this statement into a constraint on scales as follows: To say that a generalized symmetry $G^{(p)}$ emerges below some scale $E_G$ is to say that, at energies $E \ll E_G$, the theory admits a consistent coupling to a background gauge field for $G^{(p)}$, with the associated background gauge redundancy realized as a redundancy of the infrared effective description. If, however, the background gauge transformations for $G^{(p)}$ necessarily involve background fields for other symmetries, as in Equations \ref{eq:hg-maxwell-3+1d-a}--\ref{eq:hg-maxwell-3+1d-c}, then the corresponding background coupling can only be meaningful in a regime where those other symmetries are themselves already emergent. Consequently, the scale at which a given symmetry can emerge is bounded above by the scales at which the symmetries appearing in its background gauge transformations have emerged.

Applying this logic to the present case is straightforward. First, the background gauge field $C^{(3)}$ for the winding symmetry transforms with terms involving the electric 1-form background $B_e^{(2)}$ (equivalently the modified field strength $G^{(4)}$ in Equation \ref{eq:axion-Maxwell-G} contains a term proportional to $B_e^{(2)} \wedge B_e^{(2)}$). Thus, the regime in which the electric 1-form symmetry can be treated as unscreened cannot lie parametrically above the regime in which the winding symmetry background is well defined. This implies the parametric inequality
\begin{equation}
E_{\text{electric}} \lesssim E_{\text{winding}} \, .
\end{equation}
Second, the background gauge field $B_m^{(2)}$ for the magnetic 1-form symmetry transforms with terms involving both the axion-shift background $X^{(1)}$ and the electric 1-form background $B_e^{(2)}$ (equivalently the modified field strength $H^{(3)}$ in Eq.~\eqref{eq:axion-Maxwell-H} contains a term proportional to $X^{(1)} \wedge B_e^{(2)}$). As a result, the magnetic 1-form symmetry must emerge parametrically at or above the regime in which either the axion-shift symmetry or the electric 1-form symmetry admit consistent background couplings. Translating the latter condition into the absence of electric screening yields
\begin{equation}
\min\left\{E_{\text{shift}},\,E_{\text{electric}}\right\} \lesssim E_{\text{magnetic}} \, .
\end{equation}
We emphasize that these are parametric inequalities: they follow directly from the structure of the coupled background gauge redundancies and do not rely on any specific microscopic mechanism for how the symmetries are ultimately broken.

Alternatively, one can derive an analogous but more powerful set of emergence constraints from the non-invertible symmetries,
\begin{equation}
    E_{\text {electric }} \lesssim \min \left\{E_{\text {winding }}, E_{\text {magnetic }}\right\} , \quad E_{\text {shift }} \lesssim E_{\text {magnetic }},
\label{eq:3+1d-constraints}
\end{equation}
which we review in Appendix \ref{sec:non-invertible}. Ultimately, the latter is based on the fact that constructing the non-invertible symmetry defect for the axion-shift symmetry requires half-gauging the magnetic 1-form symmetry, while the former is based on the fact that constructing the non-invertible symmetry defect for the electric 1-form symmetry requires half-higher-gauging the magnetic 1-form and 2-form winding symmetries. In Section \ref{sec:interpreting-constraints}, we will focus on the constraints given in Equation \ref{eq:3+1d-constraints}, since they are stronger than those given by the higher-group.

\subsection{Interpreting the Constraints} \label{sec:interpreting-constraints}

Our main goal is to understand the phenomenological implications of the emergence constraints. To do so, we will first explore the way in which emergence constraints are satisfied in various perturbative UV completions, where the degrees of freedom responsible for breaking the relevant symmetries are automatically present in the UV. In these cases, perturbativity typically leads to a separation of scales that ensures the anomalous symmetries participating in the higher symmetry structure emerge parametrically below the non-anomalous ones. 

\subsubsection{3+1d KSVZ model} \label{sec:ksvz}

Following \cite{emergent-hg-brennan} (which explored the analogous model as a UV completion of axion-Yang Mills), perhaps the simplest completion of the axion effective field theory is the KSVZ model described by the Lagrangian
\begin{equation}
    \mathcal{L}=-\frac{1}{4} F_{\mu \nu} F^{\mu \nu}+\left|\partial_\mu \Phi\right|^2+\bar{\Psi} i \slashed D \Psi-y \Phi \bar{\Psi}_L \Psi_R+\text { h.c. }-V\left(|\Phi|^2\right),
\end{equation}
which consists of a complex scalar $\Phi$ and $N_f$ flavors of charged Dirac fermions (the flavor indices are suppressed). This Lagrangian has a global Peccei-Quinn symmetry, under which
\begin{equation}
    U(1)_\text{PQ}:\Phi \rightarrow e^{i \alpha} \Phi, \quad \Psi \rightarrow e^{-i \frac{\alpha}{2} \gamma_5} \Psi
\end{equation}
The potential $V(|\Phi|^2)=\lambda\left(|\Phi|^2-f^2 / 2\right)^2$ induces a vacuum expectation value $\langle\Phi\rangle=(f+\rho)e^{ia/f}$ for the complex scalar, which gives masses to the fermions of order $m_\Psi\sim yf$. The axion-Maxwell EFT arises upon integrating out the massive fermions and radial mode. The former is most easily accomplished by making an appropriate $U(1)_\text{PQ}$ rotation to remove the axion-fermion coupling; due to the ABJ anomaly between $U(1)_\text{PQ}$ and the $U(1)$ gauge symmetry, such a rotation generates the axion-gauge-boson coupling. From this, we see that the resulting low-energy effective field theory is of the form of Equation \ref{eq:axion-Maxwell-3+1d} with $K=N_f$.

Since the electric 1-form symmetry is explicitly broken by charged particles, it is clear that $E_\text{electric}\sim yf$, above which the massive charged fermions become dynamical. The emergence scale of $U(1)_w^{(2)}$ is more ambiguous. The authors of \cite{emergent-hg-brennan} argue for $E_{\rm winding} \sim \lambda^{1/4} f$ on the grounds that this is the energy required to access the origin in field space. The argument is as follows: roughly speaking, the energy cost to excite the radial mode to the origin in a region of volume $\ell^3$ is $E \sim (\lambda f^4) \ell^3$; at such energies the smallest resolvable length is $\ell \sim \hbar c / E$, hence the energy required to fluctuate to the origin in the smallest resolvable volume is $E \sim \lambda^{1/4} f$ in natural units. At the origin the winding current becomes ill-defined at the coordinate singularity of the axion field, and accessing the origin trivializes the effective non-trivial topology in the space of fields.  

However, while this fluctuation in a compact region creates a `hole' where the axion coordinate breaks down, this doesn't furnish a boundary for the string worldsheet; the string does not decay. For the string to decay would require such a fluctuation everywhere in space, which carries infinite energy. A highly plausible alternative is $E_{\rm winding} \sim T^{1/2} \sim f$, where $T$ is the axion string tension; at this scale there are a host of physical effects associated with axion strings becoming dynamical. Among other things, the smallest resolvable closed string loops have action $S_E \sim 1$ at this scale and pair production of string loops becomes efficient. 

In any event, given the constraint $E_\text{electric}\lesssim E_\text{winding}$, it is then interesting to ask what physical effect ensures the satisfaction of the emergence constraint in the UV completion. In fact, the emergence inequality is satisfied in this case for either interpretation of $E_{\rm winding}$: the Yukawa coupling modifies the scalar potential at one loop, and for $\lambda \ll y^4 / (16 \pi^2)$ the vacuum is destabilized. Hence, presuming that we remain in the perturbative regime, such that the unitary bound for $\lambda$ is satisfied, the emergence inequality is always consistent with the inequality for vacuum stability (see Figure \ref{fig:scales}).

It is surprising that vacuum (in)stability enters to enforce constraints associated with the anomaly structure of the theory, and one might wonder whether this remains true if additional ingredients are added to stabilize the scalar potential.  As we will see, however, there are additional effects more intrinsically related to the higher symmetry structure that saturate the emergence inequality. But first, it's interesting to explore what happens in a second UV completion.

\subsubsection{4+1d Georgi-Glashow model} \label{sec:4+1d-GG}

An alternative completion involves lifting the 3+1d axion-Maxwell EFT to a 4+1d model with a compact dimension. In particular, we consider the following action
\begin{equation}
    S=\int_{\mathcal{M}_5}\frac{1}{2 g_5^2} dC \wedge \star dC+\frac{1}{2 e_5^2} F \wedge \star F-\frac{iK}{8 \pi^2} C \wedge F \wedge F
\label{eq:axion-Maxwell-4+1d}
\end{equation}
where $F$ is the field strength of the 5d $U(1)$ gauge field $A$ that becomes the 4d Maxwell theory, while $dC$ is the field strength of the 5d $U(1)$ gauge field $C$ whose fifth component becomes the 4d axion. Here $e_4^2 \equiv e_5^2 /(2 \pi R)$, $f^2 \equiv 1 / g_5^2(2 \pi R)$ and the fifth dimension is placed on an $S^1 / \mathbb{Z}_2$ orbifold of radius $R$. The boundary conditions of the orbifold are chosen such that the zero modes of $C_\mu$ and $A_5$ are projected out \cite{Sundrum,5daxionYangMillsI,5daxionYangMillsII}. Hence, Equation \ref{eq:axion-Maxwell-4+1d} reduces to Equation \ref{eq:axion-Maxwell-3+1d} upon completing a Kaluza-Klein (KK) decomposition and integrating out the heavy Kaluza-Klein modes. 

Such a perspective provides a nice resolution to the axion quality problem. In general 3+1d axion effective field theories, $U(1)_\text{PQ}$-violating terms can shift the minimum of the axion potential away from $\theta=0$. At this level, there is no reason to forbid such terms, since $U(1)_\text{PQ}$ is both an \textit{anomalous} symmetry and a \textit{global} symmetry \cite{Reece}. However, by lifting these theories to 4+1d theories as described above, the $U(1)_\text{PQ}$ symmetry now becomes a $U(1)$ gauge symmetry, and such corrections are now controlled by higher-form symmetries. 

In our current context, this uplifting provides another UV completion with which we can study the higher symmetry emergence constraints. To begin, we note that all of the generalized symmetries of the 4+1d theory considered in Equation \ref{eq:axion-Maxwell-4+1d} have a one-to-one correspondence with those of the 3+1d theory. In particular, the zero-form axion-shift and 2-form winding symmetries map to electric 1-form and magnetic 2-form symmetries for $C$, whose conserved currents are 
\begin{equation}
    \star J_e^{(2)}=\frac{i}{g_5^2} \star d C, \quad \star J_m^{(3)}=\frac{1}{2 \pi} d C,
\end{equation}
respectively. As before, there is an electric 1-form symmetry for $A$, and the magnetic 1-form symmetry for $A$ now becomes a magnetic 2-form symmetry. In the presence of the Chern-Simons term, the electric 1-form symmetries for $C$ and $A$ are broken down to discrete non-invertible symmetries, which we review in Appendix \ref{sec:non-invertible}. This follows from the ABJ anomalies,
\begin{equation}
    d\star J_{e,C}^{(2)}=\frac{K}{8\pi^2}F\wedge F,\quad d\star J_{e,A}^{(2)}=-\frac{K}{4\pi^2}d(C\wedge F)\ ,
\end{equation}
which follow from the equations of motion. The invertible $\mathbb{Z}_K$ subgroups of these non-invertible electric 1-form symmetries are generated by the charges
\begin{equation}
    Q^{(C)}(\Sigma_3)=\int_{\Sigma_3}\star J_{e,C}^{(2)}-\frac{K}{8\pi^2}A\wedge F,\quad Q^{(A)}(\Sigma_3)=\int_{\Sigma_3}\star J_{e,A}^{(2)}+\frac{K}{4\pi^2}C\wedge F \label{eq:charges-5d}
\end{equation}
and $\alpha=2\pi n/K$ for $n\in\mathbb{Z}$ is required for invariance under large gauge transformations of $C$ and $A$. The higher-group structure also follows straightforwardly from the discussion above. In particular, upon gauging these invertible $\mathbb{Z}_K$ symmetries as well as the magnetic 2-form symmetries, Equation \ref{eq:axion-Maxwell-4+1d} becomes
\begin{align}
S&=\frac{1}{2g_5^2}\int_{\mathcal{M}_5}\left ( dC-B_{e,C}^{(2)} \right )\wedge\star \left ( dC-B_{e,C}^{(2)} \right )+\frac{i}{2\pi}\int_{\mathcal{M}_5}dB_{m,C}^{(3)} \wedge C \\
&+\frac{1}{2e_5^2}\int _{\mathcal{M}_5}\left ( F-B_{e,A}^{(2)} \right )\wedge\star \left ( F-B_{e,A}^{(2)} \right )-\frac{i}{2\pi}\int_{\mathcal{M}_5}dB_{m,A}^{(3)} \wedge A \\
&-\frac{iK}{8\pi^2}\int_{\mathcal{M}_6}\left ( dC-B_{e,C}^{(2)} \right ) \wedge \left (F - B_{e,A}^{(2)} \right ) \wedge \left (F - B_{e,A}^{(2)} \right )\ .
\end{align}
where $\partial\mathcal{M}_6=\mathcal{M}_5$, $B_{e,C}^{(2)}$ and $B_{e,A}^{(2)}$ are $\mathbb{Z}_K$ background gauge fields for the electric 1-form symmetries of $C$ and $A$, and $B_{m,C}^{(3)}$ and $B_{m,A}^{(3)}$ are background gauge fields for the magnetic 2-form symmetries of $C$ and $A$. As before, the final term is defined on a bulk 6-dimensional manifold $\mathcal{M}_6$ to cancel the mixed 't Hooft anomaly involving the electric 1-form symmetries. The higher-group structure is then revealed by noting that the physical theory on $\mathcal{M}_5$ cannot depend on the bulk extension. In particular, the field strengths for the magnetic 2-form symmetries must be modified as
\begin{align}
dB_{m,C}^{(3)}&\mapsto G^{(4)}=d B_{m,C}^{(3)}+\frac{K}{4 \pi} B_{e,A}^{(2)} \wedge B_{e,A}^{(2)} \\
dB_{m,A}^{(3)}&\mapsto H^{(4)}=d B_{m,A}^{(3)}+\frac{K}{2 \pi} B_{e,C}^{(2)} \wedge B_{e,A}^{(2)}
\end{align}
Invariance of these modified field strengths then requires the following modified set of background gauge field transformations
\begin{align}
& B_{m,C}^{(3)} \longmapsto B_{m,C}^{(3)}+d \Lambda_{m,C}^{(2)}-\frac{K}{2 \pi} B_{e,A}^{(2)} \wedge d \Lambda_{e,A}^{(1)}-\frac{K}{4 \pi} \Lambda_{e,A}^{(1)} \wedge d \Lambda_{e,A}^{(1)} \\
& B_{m,A}^{(3)} \longmapsto B_{m,A}^{(3)}+d \Lambda_{m,A}^{(2)}+\frac{K}{2 \pi} B_{e,C}^{(2)} \wedge \Lambda_{e,A}^{(1)}-\frac{K}{2 \pi} B_{e,A}^{(2)}\wedge\Lambda_{e,C}^{(1)}+\frac{K}{2 \pi} d \Lambda_{e,C}^{(1)} \wedge \Lambda_{e,A}^{(1)}
\end{align}
under $\delta B_{e,A}^{(2)}=d \Lambda_{e,A}^{(1)}$ and $\delta B_{e,C}^{(2)}= d\Lambda_{e,C}^{(1)}$. With this 3-group symmetry, the gauged action is fully invariant under all background gauge field transformations provided that the following inflow action is also present
\begin{equation}
    S_{\text {inflow }}=\frac{i}{2 \pi} \int_{\mathcal{M}_6} G^{(4)} \wedge B_{e,C}^{(2)}+\frac{i}{2 \pi} \int_{\mathcal{M}_6} H^{(4)} \wedge B_{e,A}^{(2)} .
\end{equation}
Ultimately, the emergence constraints resulting from the higher-group structure are
\begin{equation}
    E_{\text{electric}}^{(A)}\lesssim E_{\text{magnetic}}^{(C)},\quad \min\left \{ E_{\text{electric}}^{(C)},E_{\text{electric}}^{(A)} \right \} \lesssim E_{\text{magnetic}}^{(A)}
\end{equation}
where $E_{\text{magnetic}}^{(\alpha)}$ and $E_{\text{electric}}^{(\alpha)}$ are the emergence scales for the magnetic 2-form and electric 1-form symmetries for the gauge fields $\alpha=C,A$. Alternatively, one can again derive a more powerful set of emergence constraints from the non-invertible symmetries
\begin{equation}
    E_{\text{electric}}^{(A)} \lesssim \min\{E_{\text{magnetic}}^{(C)},E_{\text{magnetic}}^{(A)}\},\quad E_{\text{electric}}^{(C)} \lesssim E_{\text{magnetic}}^{(A)}\ .
\label{eq:4+1d-constraints}
\end{equation}
which we review in Appendix \ref{sec:non-invertible}. 

Again, it is natural to ask how these emergence constraints are satisfied in various UV completions. While 5d gauge theories are themselves not UV complete, we can nonetheless construct a partial UV completion for Equation \ref{eq:axion-Maxwell-4+1d} in which the degrees of freedom violating the magnetic 2-form symmetry of $C$ appear automatically. To do so, we imagine the $U(1)$ gauge field $C$ descends from a spontaneously broken $SU(2)$ gauge theory as in \cite{5dGG} \footnote{Here, we have slightly abused notation by treating the final three terms as four-forms rather than scalars.},
\begin{equation}
    S=\int_{\mathcal{M}_5}\frac{1}{2g_5^2}\text{Tr}(F_C\wedge\star F_C)+\frac{1}{2e_5^2}F_A\wedge\star F_A+\frac{1}{2}\text{Tr}(D\Phi\wedge\star D\Phi)+i\bar\Psi\slashed{D}\Psi+y\bar\Psi\Psi\Phi-V(\Phi)
\label{eq:5dGG}
\end{equation}
where $F_C$ and $F_A$ are field strengths for $SU(2)$ and $U(1)$ gauge fields $C$ and $A$, the fermion $\Psi$ lives in the fundamental of $SU(2)$ and is additionally coupled to the $U(1)$ gauge field, the Higgs field $\Phi$ lives in the adjoint of $SU(2)$, and $V(\Phi)=\lambda\left ( \Phi^a\Phi^a-v^2 \right )^2$ induces Higgsing of $SU(2)$ down to $U(1)$. It is a straightforward generalization of the calculations in \cite{Bonetti2013,Adachi:2021rjw} to show that integrating out the heavy fermions generates the desired $C\wedge F \wedge F$  Chern-Simons term.\footnote{Integrating out the heavy fermions also generates a pure Chern-Simons term of the form $C\wedge dC \wedge dC$. If we like, we can take $C$ to be the only gauge field in Equation \ref{eq:5dGG}, and place the fifth dimension on $S^1$ rather than $S^1/\mathbb{Z}_2$. Then, no zero modes are projected out by the orbifold boundary conditions, and $C\wedge dC \wedge dC$ reduces to $\theta F \wedge F$ as well as couplings to higher KK modes. Otherwise, this term does not affect our conclusions nor does it lead to any additional terms in the 3+1d theory upon neglecting terms involving heavy KK modes.} With this completion, the solitonic objects that couple to the magnetic 2-form symmetry for $C$ are 't Hooft-Polyakov monopole strings.

An advantage of the 5d UV completion for axion-Maxwell theory is that it enables us to make a slightly more precise argument for the emergence scale of the magnetic 2-form symmetry that descends into the 2-form winding symmetry of axion-Maxwell. In particular, we may compute the magnetic 2-form emergence scale according to the criterion presented in \cite{I4}, which we now review.

Recall that a $p$-form symmetry defect operator $U_\alpha$ acts on a $p$-dimensional charged operator as 
\begin{equation}
    U_\alpha\left(\Sigma^{d-p-1}\right) W_q\left(\Gamma^p\right)=e^{i q \alpha} W_q\left(\Gamma^p\right)
\end{equation}
assuming that $\Sigma$ links with $\Gamma$. In a regime where the $p$-form symmetry is broken by small effects, $U_\alpha$ is no longer topological, and the right-hand side becomes dependent on the size of $\Sigma$. If we take $\Sigma$ to be a sphere of radius $r$, which we denote as $S_r^{d-p-1}$, then one can define the following radially-dependent effective charge
\begin{equation}
    q(r)\equiv\lim _{\alpha \rightarrow 0} \frac{1}{i \alpha} \log \left(\frac{\left\langle W^{\dagger}\left(\Gamma^{p}\right)(\infty) U_\alpha\left(S_r^{d-p-1}\right) W\left(\Gamma^{p}\right)(0)\right\rangle}{\left\langle W^{\dagger}\left(\Gamma^{p}\right)(\infty) W\left(\Gamma^{p}\right)(0)\right\rangle}\right)
\end{equation}
where $W^\dagger(\infty)$ and $W(0)$ are well-separated in spacetime. The symmetry-breaking (emergence) scale may then be identified as the energy scale $\Lambda_\star\sim r_\star^{-1}$ at which 
\begin{equation}
    \left.\frac{1}{q_{\infty}}\left(\frac{d}{d \log r}\right) q(r)\right|_{r=r_{\star}} \sim O(1) .
\label{eq:sb-criterion}
\end{equation}
Following the Uehling potential calculation in QED, the authors of \cite{I4} demonstrate that this criterion is consistent with the expectation that the emergence scale of the electric 1-form symmetry is the mass of the lightest charged particle. Of the many other examples considered, the authors in \cite{I4} additionally compute the emergence scale for $U(1)_m^{(1)}$ in a 3+1d Georgi-Glashow model, and the following calculation is a straightforward generalization of this to our 4+1d completion.

To probe the symmetry-breaking effects at short distances, we must extrapolate the magnetic current operator to the UV. A current which flows to the magnetic 2-form symmetry current for $C$ in the IR is given by
\begin{equation}
    \star\mathcal{J}\equiv \frac{1}{4\pi v}\text{Tr}(\Phi F_C)\leadsto \frac{1}{2\pi}dC=\star J_m^{(3)}
\end{equation}
The effective charge $q(r)$ then approximately satisfies
\begin{equation}
    e^{i \alpha \int_{S_r^2} \star\mathcal{J}} T_\gamma^{\mathrm{UV}}\left(q_{\infty}\right)=e^{i \alpha q(r)} T_\gamma^{\mathrm{UV}}\left(q_{\infty}\right)
\label{eq:TUV}
\end{equation}
where $T_\gamma^\text{UV}(q_\infty)=\exp \left ( -\frac{4\pi^2q_\infty}{g_5^2}\int_H \mathcal{J} \right )$ and $\partial H=\Sigma_2$ is the worldsheet of a monopole string situated at $x_i=0$ for $i=1,2,3$ for all of time and extended along the orbifold direction. In the weak-coupling limit, we can expand Equation \ref{eq:TUV} to obtain an explicit expression for the effective charge
\begin{equation}
    q(r)=\left\langle\left(i \int_{S_r^2} \star\mathcal{J}\right)\left(\frac{-4 \pi^2 q_{\infty}}{g_5^2} \int_H  \mathcal{J}\right)\right\rangle+O(\lambda, g_5)
\end{equation}
Recalling that the orbifold boundary conditions are chosen such that the zero modes of $C_\mu$ are projected out, the current-current correlation function yields
\begin{equation}
    \left\langle \mathcal{J}^{A B(0)}(\vec{x}) \mathcal{J}_{A B}^{(0)}(0)\right\rangle=g_5^2\left\langle\partial_\mu \theta \partial^\mu \theta\right\rangle+\frac{g_5^2}{v^2}\langle H(x) H(0)\rangle\left\langle\partial_\mu \theta \partial^\mu \theta\right\rangle+O\left(g_5, \lambda\right)
\end{equation}
where $(0)$ denotes that we have neglected heavy KK modes and $H(x)$ describes fluctuations around the vacuum: $\Phi(x)=v+H(x)$. Ultimately, one finds
\begin{equation}
    q(r)=\frac{q_{\infty}}{g_5^2R^2} \int_0^{+\infty} d p\left(p^2 B\left(p^2\right)+C\left(p^2\right)\right)\left(r+\frac{i}{p}\right) e^{i p r}
\end{equation}
where the structure functions are given by
\begin{equation}
    B\left(p^2\right)=\frac{g_5^2}{2 \pi p^2}\left(1+\frac{p^2}{8 \pi v^2} \int_0^1 d x x^2 \log D\right), \quad C\left(p^2\right)=-\frac{g_5^2}{v^2} \frac{1}{(4 \pi)^2} \int_0^1 d x D \log D
\end{equation}
where $D \equiv x(1-x) p^2+m_H^2 x$. The final result is 
\begin{equation}
    \left|\frac{1}{q_{\infty}} \frac{d q(r)}{d \log (r)}\right| \sim \frac{1}{r^2 v^2 R}+O(g_5, \lambda),
\end{equation}
which implies that the emergence scale for the magnetic 2-form symmetry for $C$ is $\Lambda_\star \sim v\sqrt{R}$ according to the criterion in Equation \ref{eq:sb-criterion}.

\begin{figure}
    \centering
    \includegraphics[width=0.6\linewidth]{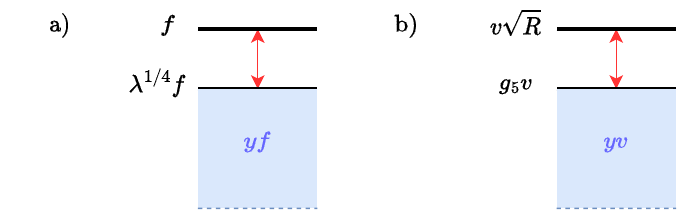}
    \caption{a) For the 3+1d KSVZ model, the vacuum stability constraint $yf\lesssim \lambda^{1/4}f$ implies that $E_\text{electric}\sim yf$ and $E_\text{winding}\sim f$ are separated whenever $\lambda$ does not saturate the unitarity bound $\lambda \lesssim 16\pi^2$. b) For the 4+1d Georgi-Glashow model, the vacuum stability constraint $yv \lesssim g_5v$ implies that the $E_\text{electric}^{(A)}\sim yv$ and $E_\text{magnetic}^{(C)}\sim v\sqrt{R}$ are separated whenever $g_5$ does not saturate the unitarity bound $g_5\lesssim \sqrt{R}$. }
    \label{fig:scales}
\end{figure}

Given the constraint $E_\text{electric}^{(A)}\lesssim E_\text{magnetic}^{(C)}$, we therefore expect that the effective field theory breaks down when the Yukawa coupling $y$ is taken to be sufficiently large compared to $\sqrt{R}$. Indeed, up to $\mathcal{O}(1)$ factors this is simply the unitarity bound of the 5d coupling. Much like in the 4d KSVZ example, however, in the perturbative regime additional effects enter before the inequality is saturated. To 1-loop order, the $\beta$-function for the quartic coupling is given by \cite{Cacciapaglia2026}
\begin{equation}
    \beta_{\tilde\lambda}(\tilde{g}_5,\tilde{y},\tilde\lambda)=\tilde{\lambda}+a_\lambda\tilde\lambda^2+b_\lambda\tilde{g}_5^4-c_\lambda\tilde{y}^4-d_\lambda\tilde{g}_5^2\tilde{\lambda}+e_\lambda\tilde{\lambda}\tilde{y}^2
\end{equation}
where $\tilde{g}_5=\mu^{1/2}g_5$, $\tilde{y}=\mu^{1/2}y$ and $\tilde{\lambda}=\mu\lambda$ are dimensionless parameters defined in terms of the energy scale $\mu$ and $a_\lambda,b_\lambda,c_\lambda,d_\lambda,e_\lambda>0$. \footnote{As described in \cite{Dienes1998,Dienes1999,Bhattacharyya2007}, the effect of summing over Kaluza-Klein modes in the 1-loop corrections to the propagators and vertex functions is to generate a power law behaviour in the running. Since the precise values of these coefficients do not affect our conclusions, we avoid explicitly computing them here.} This reveals that the vacuum destabilizes for the case of $y \gg g_5$, which is consistent with the discussion of \cite{Bucci2003}. As illustrated in Figure \ref{fig:scales}, the vacuum stability constraint $y\lesssim g_5$ is consistent with the emergence constraint $E_\text{electric}^{(A)}\lesssim E_\text{magnetic}^{(C)}$ due to the unitarity bound $g_5\lesssim\sqrt{R}$.

\subsubsection{Anomaly Inflow Mechanisms} \label{sec:inflow}

In the previous sections, we interpreted the emergence inequality between the electric symmetry for $A$ and the magnetic symmetry for the axion by considering specific UV completions. In particular, we found that the vacuum is destabilized parametrically before the emergence constraint is saturated in both the 3+1d KSVZ model and the 4+1d Georgi-Glashow model. In some sense, this is because perturbative UV completions in which stringlike objects emerge as topological defects introduce a parametric separation of scales. 

On the other hand, the emergence constraints must be satisfied {\it regardless} of the UV completion -- in particular, they should be satisfied even if we do something as simple as add a dynamical string to the theory without any further structure. Indeed, as noted in \cite{SHS-non-invertible-II}, it turns out that all of the emergence constraints from generalized symmetries are satisfied by anomaly inflow mechanisms for the topological defects, which exist \textit{regardless} of the UV completion. In this section, we review these inflow mechanisms, and provide additional arguments for why the association of inflow mechanisms with these emergence constraints is justified.

Let us focus on the 3+1d case for now. In the presence of an axion string with worldsheet $\Sigma_2$, the bulk topological term lacks invariance under 0-form gauge transformations of $A$ \cite{Callan-Harvey,Naculich,harvey-anomalies}. In particular, under $\delta_\Lambda A=d\Lambda$, one has
\begin{equation}
    \delta_\Lambda S=-\frac{iK}{8\pi^2f}\int_{\mathcal{M}_4}d^2a\wedge \Lambda F=\frac{iK}{4\pi}\int_{\Sigma_2} \Lambda F
\label{eq:3+1d-axion-string-anomaly}
\end{equation}
where we have used the fact that $d^2a=2\pi\delta(\Sigma_2)$. Similarly, in the presence of a magnetic monopole with worldline $\Sigma_1$, the bulk topological term lacks invariance under ($-1$)-form gauge transformations of $a$. In particular, under $\delta_\lambda a=2\pi\lambda f$ for $\lambda\in\mathbb{Z}$, one has
\begin{equation}
    \delta_\lambda S=-\frac{iK}{4\pi}\int_{\mathcal{M}_4}\lambda A\wedge dF=iK\int_{\Sigma_1} \lambda A
\label{eq:3+1d-monopole-anomaly}
\end{equation}
where we have used the fact that $dF=4\pi\delta(\Sigma_1)$ for a magnetic monopole. Gauge invariance then requires that axion strings and magnetic monopoles host anomalous degrees of freedom that exactly cancel the bulk anomalies given in Equations \ref{eq:3+1d-axion-string-anomaly} and \ref{eq:3+1d-monopole-anomaly}. Possible worldvolume actions for the axion string include one of massless charged chiral fermions, or one of massless charged bosons \cite{Naculich,Heidenreich2021}:
\begin{equation}
    S_{\text{string}}=\int_{\Sigma_2}-\frac{1}{2}d_A\phi\wedge\star d_A\phi+\frac{iK}{4\pi}\phi F
\label{eq:string-maxwell}
\end{equation}
where $d_A\phi=d\phi-A$ and $\phi$ is a compact scalar field. This bosonic theory is related to the fermionic worldvolume action by bosonization \cite{Naculich}, and has the advantage of translating one-loop effects in the fermionic theory into classical effects. In particular, Equation \ref{eq:string-maxwell} precisely cancels the bulk anomaly under the gauge variation $\delta A= d\Lambda$, $\delta\phi=\Lambda$. Similarly, a possible worldvolume action for the magnetic monopole is \cite{Jackiw1975,MonopoleLoops,SHS-non-invertible-II,Fukuda2021}
\begin{equation}
    S_\text{monopole}=\int_{\Sigma_1}-\frac{1}{2}l_\sigma {d}_A \sigma \wedge \star {d}_A \sigma-\frac{iK}{2 \pi f}a d_A \sigma
\label{eq:monopole-action-3+1d}
\end{equation}
where $\sigma$ is again a compact scalar field. As we will review briefly, the electric charge resulting from this inflow is precisely given by $Q=-K\theta/2\pi$ \cite{Fukuda2021}, which is the well-known Witten effect. These observations are consistent with $E_{\text {electric}} \lesssim \min \left\{E_{\text {winding }}, E_{\text {magnetic }}\right\}$. Namely, the presence of charged zero modes on axion strings and magnetic monopoles implies that such defects break the electric 1-form symmetry. Furthermore, it was recently shown that loops of magnetic monopoles generate an axion potential due to the Witten effect \cite{MonopoleLoops}, which is consistent with $E_{\text {shift }} \lesssim E_{\text {magnetic }}$. In particular, the topological term in the worldline action for the magnetic monopole, which describes a quantum particle on a circle \cite{Gaiotto2017}, implies a $\theta$-dependent mass spectrum for the dyons, which leads to $\theta$-dependence in the effective potential generated by monopole loops.  

Given that both the generalized (higher-group and non-invertible) symmetries and the inflow mechanisms are generated by the axion-gauge-boson coupling, it perhaps is not surprising that these inflow mechanisms satisfy all of the emergence constraints. The connection can be made even more explicit by noting that $Q(\Sigma_3)$ and $Q(\Sigma_2)$ (Equation \ref{eq:charges}) have the interpretation of magnetic brane couplings when defined on open manifolds (possess a boundary). In particular, the former describes the worldsheet of an axion string on $\partial\Sigma_3$ while the latter describes the worldline of a magnetic monopole on $\partial\Sigma_2$. $\star J^{(1)} \sim\star da$ and $\star J_e^{(2)} \sim \star dA$ are the standard magnetic brane couplings for axion string and magnetic monopoles, while the additional terms are required to ensure that the couplings do not depend on the choice of auxiliary surfaces $\Sigma_3$ and $\Sigma_2$. Similar to the story for the symmetry defects, such couplings are topological but now lack gauge-invariance. Hence, the axion string and magnetic monopole defects must carry anomalous degrees of freedom by anomaly inflow. \footnote{For example, as described in \cite{Wen_QFT,QCD-SPT}, under a $U(1)$ gauge transformation $\delta A = d\Lambda$, the bulk Chern-Simons term $S=\frac{iK}{4\pi}\int_{\mathcal{M}_3}A\wedge dA$ transforms as $\delta S=\frac{iK}{4\pi}\int_{\mathcal{M}_2}\Lambda dA$ in the presence of a boundary $\mathcal{M}_2=\partial\mathcal{M}_3$. Hence, $\mathcal{M}_2$ must carry massless fermions or bosons which couple to $A$ such that the resulting boundary anomaly cancels the bulk anomaly. The discussion above recasts the anomaly inflow for axion strings in terms of the usual anomaly inflow between a bulk topological phase and its boundary.} For the case of the magnetic monopole, this explicitly describes the Witten effect.\footnote{As described in \cite{Fukuda2021} if the bulk and boundary currents on these auxiliary surfaces are $J_\text{bulk}$ and $J_\text{bdry}$, then the total current $\mathcal{J}=J_\text{bulk}+J_\text{bdry}$, which we formally define as $\delta S=(2\pi)^{-1}\int_{\Sigma}\delta A\wedge \mathcal{J}$ for a defect defined on worldvolume $\partial\Sigma$, must be conserved by gauge-invariance: $d\mathcal{J}=0$. For the case of the axion string and magnetic monopole, it follows that the boundary currents are given by
\begin{equation}
     J_\text{bdry}^{(s)}=\frac{K}{4\pi}A,\quad J_\text{bdry}^{(m)}=-\frac{K}{2\pi f}a,
\end{equation}
respectively. The associated charges are given by 
\begin{equation}
    Q^{(s)}(\Sigma_1)=\frac{K}{4\pi}\int_{\Sigma_1}A,\quad Q^{(m)}(\Sigma_0)=-\frac{K}{2\pi f }a(\Sigma_0)
\end{equation}
where $\Sigma_1$ and $\Sigma_0$ describe an axion string and magnetic monopole at fixed time. The latter of course is the familiar Witten effect. This shows explicitly that the Witten effect is a consequence of anomaly inflow.}

In fact, the axion-shift and electric symmetry-breaking effects of the topological defects can be derived explicitly from the magnetic brane couplings and the observation that 't Hooft anomalous symmetries end at the boundary \cite{Jensen2018,Thorngren2021}. Namely, the modification to $Q(\Sigma_3;\partial\Sigma_3)$ due to the axion-gauge-boson coupling implies that the worldvolume for an axion-string is attached to a 2+1d pure Chern-Simons theory. As reviewed in Appendix \ref{sec:FQH}, such a theory has a 't Hooft anomaly for the $\mathbb{Z}_K$ electric 1-form symmetry. Hence, axion strings break this electric 1-form symmetry, which is consistent with the inflow-induced charged zero modes. Similarly, the modification to $Q(\Sigma_2;\partial\Sigma_2)$ implies that the worldvolume for a magnetic monopole is attached to a 1+1d BF theory. As reviewed in Appendix \ref{sec:ZNGT}, such a theory has a mixed 't Hooft anomaly for the $\mathbb{Z}_K$ axion-shift and electric 1-form symmetries. Once again, this is consistent with both the inflow-induced charged zero modes and the observation that loops of magnetic monopoles generate an axion potential.

\begin{figure}
    \centering
    \includegraphics[width=0.3\linewidth]{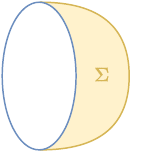}
    \caption{The magnetic brane worldvolume (blue circle) must be attached to a TQFT defined on $\Sigma$ in order for the coupling to be well-defined. If the TQFT possesses a 't Hooft anomaly for the global symmetry group $G$, then it follows that $G$ is broken on the worldvolume. In the examples in this paper, we generally find that this symmetry-breaking is consistent with the charged degrees of freedom living on the worldvolume and induced by inflow.}
    \label{fig:brane-coupling}
\end{figure}

We have demonstrated that the inflow-induced symmetry-breaking effects can be derived directly from the generalized symmetry charges, using the implication that the topological defects are attached to topological phases. Interestingly, the inflow SPTs for these topological phases appear in the modified field strengths given in Equations \ref{eq:axion-Maxwell-G} and \ref{eq:axion-Maxwell-H} (see Appendix \ref{sec:TO} for a review of the inflow actions). It was also demonstrated in \cite{Hidaka2020} that the physically observable consequences of the anomaly inflow, namely the Witten effect and the anomalous Hall effect of the axion string, may be derived from the correlation functions of the symmetry operators, and in this manner may also be understood as a direct consequence of the 3-group.

Before proceeding, we note that all of this discussion generalizes trivially to the 4+1d case. A monopole string for $C$ with worldsheet $\Sigma_2$ generates an anomalous variation of the form given in Equation \ref{eq:3+1d-axion-string-anomaly} under a gauge transformation $\delta_{\Lambda}A=d\Lambda$. The appropriate worldvolume action for the monopole string is of the form of Equation \ref{eq:string-maxwell}. Similarly, a monopole string for $A$ with worldsheet $\Sigma'_2$ generates an anomalous variation $    \delta_\Lambda S=-\frac{iK}{2\pi}\int_{\Sigma_2}d\Lambda\wedge A$ under a gauge transformation $\delta_{\Lambda}C=d\Lambda$. Hence, gauge invariance requires that both monopole strings for both $C$ and $A$ host zero modes charged under $A$, which is consistent with $E_\text{electric}^{(A)} \lesssim \min\{E_\text{magnetic}^{(C)},E_\text{magnetic}^{(A)}\}$. An appropriate worldvolume action for the monopole string for $A$ is given by
\begin{equation}
    S_{\text {monopole }}=\int_{\Sigma_2}-\frac{1}{2} \frac{T_M}{m_W^2} d_A \sigma \wedge \star d_A \sigma-\frac{i K}{2 \pi} C\wedge d_A \sigma
\label{eq:dirac-string-action}
\end{equation}
where $m_W$ is the $W$ boson mass and $T_M$ is the monopole string tension. As described in \cite{craig2024highqualityaxionshigherformsymmetries}, loops of such monopole strings generate an axion potential, which may be computed by completing a KK decomposition of \ref{eq:dirac-string-action} and following the calculation in \cite{MonopoleLoops} for the resulting 3+1d monopole worldline action (terms involving higher KK modes may be neglected since they do not couple to the axion). Ultimately, the generation of an axion potential via loops of monopole strings is consistent with $E_\text{electric}^{(C)} \lesssim E_\text{magnetic}^{(A)}$. In fact, for a modification to the 4+1d completion considered in Section \ref{sec:4+1d-GG}, whereby the gauge field $A$ is taken to be the $SU(2)$ gauge field as opposed to $C$, $E_\text{magnetic}^{(A)}$ can be explicitly computed according to criterion given in Equation \ref{eq:sb-criterion}. One finds $E_\text{magnetic}\sim v\sqrt{R}$, which is consistent with the emergence constraint and $E_\text{electric}^{(C)}\sim v\sqrt{R}$, where the latter follows from the coefficient on the axion potential generated by monopole loops \cite{craig2024highqualityaxionshigherformsymmetries}.

\section{Axion Yang-Mills} \label{sec:aym}

We turn next to the emergence constraints of axion Yang-Mills with gauge-group $G$, for which the action is given by
\begin{equation}
    S=\int_{\mathcal{M}_4}\frac{1}{2}da\wedge\star da+\frac{1}{2e_4^2}\text{tr}\left(F\wedge\star F\right)-\frac{iK}{8\pi^2f}a \text{tr}\left (F\wedge F \right )
\label{eq:3+1d-axion-YM}
\end{equation}
where $F$ is now a non-abelian field strength for a $G$ gauge field. We first discuss the case of $G=SU(N)$ and then discuss the more general case of $G=SU(N)/\mathbb{Z}_p$.

\subsection{$G=SU(N)$} \label{sec:SU(N)}

$SU(N)$ Yang-Mills possesses a $\mathbb{Z}_N^{(1)}$ 1-form center symmetry, which plays the role of $U(1)_e^{(1)}$ in a $U(1)$ gauge theory. In particular, under a center-symmetry transformation $\delta A=A_{\mathbb{Z}_N}$ by a discrete $\mathbb{Z}_N$ gauge field $A_{\mathbb{Z}_N}$, the Wilson line transforms as
\begin{equation}
    W_R(\gamma)=\operatorname{tr}_R \exp \left[i \oint_\gamma A\right] \rightarrow z^{I_N(R)} W_R(\gamma)
\end{equation}
where $I_N(R)$ is the $N$-ality of the representation and $z=e^{i\oint_\gamma A_{\mathbb{Z}_N}}\in\mathbb{Z}_N$ since $A$ is a discrete $\mathbb{Z}_N$ gauge field. On the other hand, there is no magnetic 1-form symmetry \cite{Olive1978,Tong2017, brennan-hong}, and hence, it is unclear if non-invertible axion-shift symmetry defects exist. However, one can construct a 3-group following the procedure for gauging $\mathbb{Z}_N^{(1)}$  described in \cite{Gaiotto2017,anomalies-coupling-constants,brennan-hong,tonggt}. To gauge $\mathbb{Z}_N^{(1)}$, we begin by lifting the $SU(N)$ gauge field to a $U(N)$ gauge field using the appropriate Lagrange multiplier constraint,\begin{equation}
    S=\int_{\mathcal{M}_4}\frac{1}{2}da\wedge\star da+\frac{1}{2e_4^2}\text{tr}\left(\hat{F}\wedge\star \hat{F}\right)-\frac{iK}{8\pi^2f}a \text{tr}\left (\hat{F}\wedge \hat{F} \right )+\frac{1}{2\pi}\varphi_2\wedge\text{tr}(\hat{F}).
\end{equation}
Here, $\hat{F}$ is now a $U(N)$ field strength and $\varphi_2$ is a 2-form Lagrange multiplier that ensures only the traceless part of $\hat{F}$ is dynamical. A center symmetry transformation now corresponds to shifts by a $U(1)$ gauge field
\begin{equation}
    \hat{A}\rightarrow\hat{A}+\hat{A}_{\mathbb{Z}_N}\Rightarrow \hat{F}\rightarrow\hat{F}+\hat{F}_{\mathbb{Z}_N}
\end{equation}
where $\hat{F}$ is the corresponding $U(1)$ field strength. However, there is an ambiguity in the choice of lift from ${A}_{\mathbb{Z}_N}$ to $\hat{A}_{\mathbb{Z}_N}$, which we account for by introducing a 2-form background gauge field $B^{(2)}$. Hence, the appropriate action is
\begin{align}
S & =\frac{1}{2} \int_{\mathcal{M}_4}\left(d a-f A^{(1)}\right) \wedge \star\left(d a-f A^{(1)}\right)+\int_{\mathcal{M}_4} \frac{\varphi}{2 \pi}\left(\operatorname{Tr}(\hat{F})-N B^{(2)}\right) \\
&+\frac{i}{2 \pi f} \int_{\mathcal{M}_4} a dC^{(3)} +\frac{1}{e_4^2} \int_{\mathcal{M}_4} \operatorname{Tr}\left[\left(\hat{F}-B^{(2)} \mathbf{1}\right) \wedge \star\left(\hat{F}-B^{(2)} \mathbf{1}\right)\right] \\
&-\frac{i K}{8 \pi^2 f} \int_{\mathcal{M}_5}\left(d a-f A^{(1)}\right)\operatorname{Tr}\left[\left(\hat{F}-B^{(2)} \mathbf{1}\right) \wedge\left(\hat{F}-B^{(2)} \mathbf{1}\right)\right]
\end{align}
where we have additionally gauged the axion-shift and 2-form winding symmetries in the usual manner. Again, we have defined the topological term on a five-dimensional manifold $\mathcal{M}_5$ to account for a mixed 't Hooft anomaly involving the axion-periodicity. The mixed 't Hooft anomaly results from the fact that, by gauging the center symmetry, the $\theta$ periodicity becomes $\theta\sim\theta+2\pi N$ rather than $\theta\sim\theta+2\pi$. \footnote{This holds for spin manifolds as well as non-spin manifolds with odd $N$. For non-spin manifolds and even $N$, the $\theta$-periodicity is $\theta\sim\theta+4\pi N$} In particular, upon gauging the center symmetry, the instanton number becomes
\begin{align}
\frac{1}{8\pi^2}\int\text{tr}(F\wedge F)&\mapsto\frac{1}{8\pi^2}\int\left(\operatorname{tr}[\hat{F} \wedge \hat{F}]-N {B}^{(2)} \wedge {B}^{(2)}\right) \\
&=n +N(N-1)\int\frac{B^{(2)}\wedge B^{(2)}}{8\pi^2}
\end{align}
where we have used the Lagrange multiplier constraint $\text{tr}(\hat{F})=NB^{(2)}$ and the appropriate integral class (second Chern class) for a $U(N)$ field strength
\begin{equation}
    n=\frac{1}{8 \pi^2} \int[\operatorname{tr}(\hat{F} \wedge \hat{F})-\operatorname{tr}(\hat{F}) \wedge \operatorname{tr}(\hat{F})] \, .
\end{equation}
As a $\mathbb{Z}_N$ gauge field, we have that $N \oint \frac{B_2}{2 \pi} \in \mathbb{Z}$. Hence, the minimal fractional instanton number is $1/N$. It is common to define the integral class $w_2=N \frac{B_2}{2 \pi} \in H^2\left(\mathcal{M}_4 ; \mathbb{Z}_N\right)$, in which case the second term is written as $\frac{N-1}{2N}\int w_2\wedge w_2$ \footnote{Formally, it is expressed as $\frac{N-1}{2N}\int \mathcal{P}(w_2)$ where $\mathcal{P}(w_2)$ is Pontryagin square operation. See \cite{brennan-hong,anomalies-coupling-constants,Gaiotto2017} for reviews.}, to emphasize the fractionality. Ultimately, similar to the story for axion-Maxwell, the presence of the 't Hooft anomaly resulting from this fractional instanton charge gives rise to a higher-group structure. Namely, we can apply the same bulk-boundary argument to argue that the field strength for $C^{(3)}$ must be modified. We require that
\begin{equation}
    \int_{\mathcal{M}_5 \cup_{\mathcal{M}_4} \overline{\mathcal{M}}_5^{\prime}}\left[\left(dC^{(3)}-\frac{KN(N-1)}{4 \pi} B^{(2)} \wedge B^{(2)}\right) \wedge d a\right]=0\bmod2 \pi i \mathbb{Z}
\end{equation}
and hence that \footnote{Without the bulk extension, this modification actually directly cancels the 't Hooft anomaly. Nonetheless, both arguments lead to the same conclusion. }
\begin{equation}
    dC^{(3)}\mapsto G^{(4)}=d C^{(3)}+\frac{K N(N-1)}{4 \pi} B^{(2)} \wedge B^{(2)}
\end{equation}
Invariance of this field strength requires 
\begin{align}
    C^{(3)} &\longmapsto C^{(3)}+d \Lambda^{(2)}-\frac{K N(N-1)}{2 \pi} \Lambda^{(1)} \wedge B^{(2)}-\frac{K N(N-1)}{4 \pi} \Lambda^{(1)} \wedge d \Lambda^{(1)} \\
    B_e^{(2)} &\longmapsto B_e^{(2)} + d\Lambda_e^{(1)}
\end{align}
which establishes a higher-group symmetry involving the 2-form winding symmetry and the 1-form center symmetry. Finally, invariance under all background gauge field transformations requires the inflow action,
\begin{equation}
    S_\text{inflow}=\frac{i}{2\pi}\int_{\mathcal{M}_5}G^{(4)}\wedge A^{(1)}
\end{equation}
Given that the higher-group only mixes the 1-form center and 2-form winding symmetries, the emergence constraint that follows is simply
\begin{equation}
    E_\text{center} \lesssim E_\text{winding} \, .
\label{eq:3+1d-YM-constraint}
\end{equation}
As was the case for axion-Maxwell, this constraint is consistent with anomaly inflow for axion strings. Namely, the existence of charged zero modes localized on axion strings implies that the 1-form center symmetry is broken at the scale at which axion strings become dynamical. Similar to above, this can be seen at the level of the invertible symmetry operator for the axion-shift symmetry, which is given by $\mathcal{U}_\alpha(\Sigma_3)=\exp(i\alpha Q(\Sigma_3))$ where
\begin{equation}
    Q(\Sigma_3)=\int_{\Sigma_3}\star J^{(1)}-\frac{K}{2\pi}S_\text{CS}[A;\Sigma_3]
\end{equation}
and $S_\text{CS}[A;\Sigma_3]$ is the non-Abelian Chern-Simons action for the $SU(N)$ gauge field $A$,
\begin{equation}
    S_\text{CS}[A;\Sigma_3]=\frac{1}{4\pi}\int_{\Sigma_3}\text{tr}\left ( A\wedge dA-\frac{2i}{3}A\wedge A\wedge A \right )\ .
\end{equation}
As above, for $\Sigma_3$ with a non-trivial boundary, $\star J^{(1)}$ is the standard magnetic brane coupling for axion strings, and the Chern-Simons action ensures that the axion string coupling is topological. Since the Chern-Simons action lacks gauge-invariance on the boundary, the axion string must host anomalous degrees of freedom that couple to the $SU(N)$ gauge field. 

Before proceeding, we remark that the mixed 't Hooft anomaly giving rise to the higher-group structure does not hold when $K=0\bmod N$. In this case, it is possible to cancel the bulk gauge anomaly by placing the appropriate number of flavors of fermions living in the adjoint representation of the $SU(N)$ gauge group on axion strings, since the Dynkin index $I(R)$ of such a representation is $N$, assuming standard convention $I(F)=1/2$ for the fundamental representation $F$. Then, the emergence constraint given in Equation \ref{eq:3+1d-YM-constraint} would no longer hold, since fermions living in a representation of trivial $N$-ality do not break the center symmetry. Note, however, that the bulk anomaly can also be eliminated by placing $K$ times as many flavors of fundamental fermions on axion strings, for which the center symmetry is broken. Since the only requirement for the anomalous degrees of freedom living on the axion string is that they cancel the bulk anomaly, we simply make the observation that the lack of a higher-group structure appears to imply that the emergence inequality \textit{may} be broken.\footnote{A similar discussion holds for KSVZ fermions. In the KSVZ model, $K=\sum_i 2I(R_i)$ where $i$ labels different flavors of KSVZ fermions and $R_i$ is the corresponding representation. For a single flavor of fermion living in a representation $R$ of trivial $N$-ality, $K=2I(R)$ must be divisible by $N$. As argued in \cite{ReeceTwist}, this follows since such a representation is also be a valid representation of $SU(N)/\mathbb{Z}_N$, which, as we just reviewed, has fractional instanton numbers for which the fractional part is a multiple of $1/N$. Hence, for the case of $K=0\bmod N$, valid UV theories include both ones of fermions living strictly in representations of trivial $N$-alities and ones of fermions living in representations of non-trivial $N$-alities.}

\subsection{$G=SU(N)/\mathbb{Z}_p$} \label{sec:SU(N)Zp}

We now consider the more general case of $G=SU(N)/\mathbb{Z}_p$. In this case, there is a $\mathbb{Z}_{N/p}^{(1)}$ electric 1-form symmetry and a $\mathbb{Z}_p^{(1)}$ magnetic 1-form symmetry.\footnote{More generally, if $\mathbb{Z}_M\subseteq \mathbb{Z}_N$ is the subgroup of the center which acts trivially on all of the matter fields, then the electric symmetry is $\mathbb{Z}_{M/p}^{(1)}$} The former implies that there is a higher-group presuming that $p\neq N$. Once again, the resulting emergence constraint $E_\text{center}\lesssim E_\text{winding}$ is consistent with anomaly inflow for axion strings.

Furthermore, the axion-shift symmetry is broken down to a $\mathbb{Z}_{K/p}^{(0)}$ symmetry, which follows from the fact that the instanton number in an $SU(N)/\mathbb{Z}_p$ gauge theory can in general take a minimal fractional value of $1/p$. In particular, we have \footnote{Again, formally this is expressed as $\frac{1}{8 \pi^2} \int_\mathcal{M} \operatorname{tr}(F \wedge F)=\frac{N(N-1)}{2 p^2} \int_\mathcal{M}\mathcal{P}(w_2) \bmod 1$.}
\begin{equation}
    \frac{1}{8\pi^2}\int_{\mathcal{M}}\text{tr}(F\wedge F)=\frac{N(N-1)}{2p^2}\int_{\mathcal{M}}w_2\wedge w_2\bmod 1
\end{equation}
where the second Stiefel-Whitney class $w_2\in H^2(\mathcal{M},\mathbb{Z}_p)$ labels the discrete magnetic fluxes of the $SU(N)/\mathbb{Z}_p$ gauge theory. This is a simple generalization of the calculation reviewed in Section \ref{sec:SU(N)} for a $PSU(N)$ gauge theory. In particular, we gauge the $\mathbb{Z}_p^{(1)}$ center symmetry with a discrete $\mathbb{Z}_p$ gauge field $B^{(2)}$, and define the integral class $w_2=p\frac{B^{(2)}}{2\pi}$.

For $p\neq1$, since a magnetic 1-form symmetry does exist, one may anticipate that it is possible to construct non-invertible symmetry defects for the axion-shift symmetry. Indeed, there is a non-invertible $\mathbb{Z}_K^{(0)}$ axion-shift symmetry, within which the invertible $\mathbb{Z}_{K/p}^{(0)}$ shift symmetry is a subgroup, which is generated by the following defect \cite{Cordova-non-invertible,Cordova2024-II}
\begin{equation*}
        \mathcal{D}_{{2\pi}/{K}}^{(0)}(\Sigma_3)=\int \left . Dc \right|_{\Sigma_3}\exp \left (i\oint_{\Sigma_3}\frac{2\pi}{K}\star J_s^{(1)}+\frac{p}{4\pi}c\wedge dc+\frac{1}{2\pi}c\wedge w_2 \right )
    \end{equation*}
which satisfies the non-invertible fusion rule 
\begin{equation}
    \mathcal{D}_{2\pi/K}(\Sigma_3)\times\mathcal{D}_{2\pi/K}^\dagger(\Sigma_3)\sim\sum_{\Sigma\in H_2(\Sigma_3,\mathbb{Z}_p)}\exp \left(\frac{2 \pi i}{p} \int_{\Sigma} w_2\right) .
\end{equation}
Namely, the fusion $\mathcal{D}\times\mathcal{D^\dagger}$ yields a sum over magnetic symmetry defects. Hence, one expects that the axion-shift symmetry is violated by dynamical monopoles. As we will review below, the monopole worldline action has an axion-dependent term, from which it is a straightforward generalization of \cite{MonopoleLoops} to show that loops of $SU(N)/\mathbb{Z}_p$ dyons generate an axion potential. Hence, the resulting constraint $E_\text{shift}\lesssim E_\text{mag}$ is again satisfied regardless of the UV completion. 

We may explicitly derive the worldline action for the monopole using a generalization of the discussion in Section \ref{sec:inflow} regarding magnetic brane couplings. Note that we will focus on odd $p$ to avoid subtleties with the Pontryagin square operation.\footnote{For odd $p$, the Pontryagin square operation maps $w_2$ to $w_2\cup w_2$ where $\cup$ is the standard cup product. For even $p$, it maps $w_2$ to $\tilde{w}_2 \cup \tilde{w}_2-\tilde{w}_2 \cup_1 \delta \tilde{w}_2$ where $\delta:C^i\rightarrow C^{i+1}$ is the coboundary operator, $\cup_i: C^p \times C^q \rightarrow C^{p+q-i}$ is the higher-cup product and $\tilde{w}_2$ is the integer lift of $w_2$. See \cite{Whitehead1949,Benini2019} for more details.} Following \cite{Chern-Weil-II}, we couple $w_2$ to a Lagrange multiplier field $\widetilde{B}\in H^2(\mathcal{M},\mathbb{Z}_p)$, which identifies $w_2$ with $B^{(2)}\in H^2(\mathcal{M},\mathbb{Z}_p)$. As a result, the terms in the action that possess $w_2$ are given by
\begin{equation}
    S \supset \frac{2\pi i}{p}\int \widetilde{B}\wedge w_2+ia\frac{KN(N-1)}{p^2f}\int \frac{w_2\wedge w_2}{2}
\end{equation}
Hence, in summing over $w_2$ and thereby promoting the $SU(N)$ gauge theory to an $SU(N)/\mathbb{Z}_p$ gauge theory, we find that on-shell configurations satisfy
\begin{equation}
    \oint \left ( \widetilde{B}+\frac{KN(N-1)}{2\pi pf}aw_2 \right )=0\bmod p
\end{equation}
This implies that the genuine line operator for a monopole with worldline $\gamma$ is no longer $H_k[\gamma]=e^{\frac{2\pi i k}{p}\int_D \widetilde{B}}$ for some auxiliary surface $D$ such that $\partial D=\gamma$ \footnote{Generally, $d w_2=j_\text{mag}$ in the presence of a magnetic monopole current $j_\text{mag}$. Hence, if a monopole with worldline $\gamma$ is inserted, then $w_2=\delta^{(2)}(D)$ for some auxiliary surface $D$ satisfying $\partial D=\gamma$. Without the axion-gauge-boson coupling, it follows that the magnetic line operator is simply $H_k[\gamma]=e^{\frac{2\pi i k}{p}\int_D \widetilde{B}}$}, but rather 
\begin{equation}
    H_k'[\gamma]=e^{\frac{2\pi i k}{p}\int_D \left ( \widetilde{B}+\frac{ KN(N-1)}{2\pi pf}aw_2 \right )}
\end{equation}
Analogous to the abelian case, the mixed 't Hooft anomaly generated by $\theta w_2$ implies that the shift and electric symmetries are broken on the monopole worldline. This anomaly can perhaps be made more explicit by noting that, as described in \cite{Chern-Weil-II}, one may lift the characteristic class $w_2$ to a $U(1)$ field strength subject to the gauge invariance $w_2\rightarrow w_2+p\Lambda_2$ for $\Lambda_2\in H^2(\mathcal{M},\mathbb{Z})$. Since $w_2$ is closed, we may locally write it as $w_2=\frac{1}{2\pi}dA_e$. The mixed 't Hooft anomaly then takes the precise form described in Appendix \ref{sec:ZNGT} with $A_e$ playing the role of $a^{(1)}$ and $\theta$ playing the role of $b^{(D-2)}$.

As in the abelian case, we can explicitly derive the worldline action and verify these symmetry-breaking effects. Namely, the 1+1d BF term $\theta w_2$ implies that the monopole defect must carry anomalous degrees of freedom by anomaly inflow. The appropriate gauge-invariant line operator is given by
\begin{equation}
    \widetilde{H}_k^{\prime}[\gamma]=H_k^{\prime}[\gamma] e^{\frac{i k KN(N-1)}{p^2} \int_\gamma \frac{a}{2 \pi f}\left(d \sigma-A_e\right)}
\end{equation}
where $\sigma$ is a charged scalar that describes a dyon mode on the worldline. With the explicit form of the worldline action, it is clear that loops of these monopoles generate an axion potential \cite{MonopoleLoops}. Ultimately, as is consistent with discussion so far, we find that monopoles simultaneously break both the shift and electric symmetries due to anomaly inflow.

\section{Adding charged matter} \label{sec:matter}

Thus far, we have focused on pure axion-gauge EFTs, whereas phenomenologically relevant axion models also have matter charged under the gauge group. As discussed in \cite{emergent-hg-brennan}, while the addition of charged matter explicitly breaks the electric/center 1-form symmetries, higher-group structures still persist due to resulting zero-form flavor symmetries. 

\subsection{Axion-QED}

Let us first review the case of axion electrodynamics with charged matter (axion-QED), for which the Lagrangian is
\begin{equation}
    S=\int_{\mathcal{M}_4}\frac{1}{2} d a \wedge \star d a+\frac{1}{2 g^2} F \wedge \star F-\frac{i K}{8 \pi^2 f}  a F \wedge F+ i \bar{\Psi} \slashed{D} \Psi+i \overline{\tilde{\Psi}} \slashed{D} \tilde{\Psi}+(m \tilde{\Psi} \Psi+\text{c.c.})
\label{eq:axion-QED}
\end{equation}
where $\Psi$ and $\tilde\Psi$ are Weyl fermions of opposite unit electric charge. Assuming that there are $N_f$ flavors of such Weyl fermions, the electric 1-form symmetry is replaced by the following zero-form global flavor symmetry 
\begin{equation}
    \frac{SU(N_f)^{(0)}}{\mathbb{Z}_L}\quad,\quad L=\text{gcd}(K,N_f)
\label{eq:axion-QED-flavor-symmetry}
\end{equation}
where the $\mathbb{Z}_L$ quotient indicates that all gauge-invariant operators are acted upon trivially by $\mathbb{Z}_L\subset SU(N_f)$. It is easy to see all operators constructed solely out of fermions are acted upon trivially by $\mathbb{Z}_{N_f}\subset SU(N_f)$, since such operators require an equal number of $\Psi$ and $\bar\Psi$ fermions. As discussed in \cite{emergent-hg-brennan}, one can also consider monopole operators embedded on domain walls across which the axion field shifts as $a\rightarrow a+2\pi f$. Such monopole operators carry charge $-K$ due to the presence of a level-$K$ Chern-Simons term along the wall (see Appendix \ref{sec:TO} for a review of 2+1d Chern-Simons theories). Hence, they must be dressed with $K$ modes of $\Psi$ in order to be gauge-invariant. While these dressed monopole operators have charge $K$ under $\mathbb{Z}_{N_f}$, any subgroup $\mathbb{Z}_L\subset\mathbb{Z}_{N_f}$ acts trivially on such operators assuming $K$ is a multiple of $L$. Equation \ref{eq:axion-QED-flavor-symmetry} then follows from the largest such $L$ being $\text{gcd}(K,N_f)$. Ultimately, the non-trivial quotient $\mathbb{Z}_L$, which is also a valid subgroup of the $U(1)$ gauge group, permits fractional $U(1)$ fluxes. To satisfy Dirac quantization for a matter field living in $(\mathbf{N}_f,1)$, such fluxes must be related to the discrete magnetic fluxes of the $SU(N)_f/\mathbb{Z}_L$ flavor group, which we label with $w_2\in H^2(\mathcal{M},\mathbb{Z}_L)$, as
\begin{equation}
    \int_{\Sigma_2} \frac{F}{2 \pi}=\frac{1}{L} \int_{\Sigma_2} w_2+s, \quad s \in \mathbb{Z}
\label{eq:SW-axion-QED}
\end{equation}
Then, defining $B_e^{(2)}\equiv 2\pi w_2/L$, we find that cancellation of the associated 't Hooft anomaly requires the same modified field strengths as presented in Equations \ref{eq:axion-Maxwell-G} and \ref{eq:axion-Maxwell-H}. Hence, below the scale at which operators charged under $\mathbb{Z}_L\subset SU(N_f)$ appear, there exists a higher-group structure involving the 2-form winding symmetry and the flavor symmetry, assuming $L=\text{gcd}(K,N_f)>1$. The resulting emergence constraints are
\begin{equation}
    E_{\mathbb{Z}_L} \lesssim E_{\text {winding }}, \quad \min \left\{E_{\text {shift }}, E_{\mathbb{Z}_L}\right\} \lesssim E_{\text {magnetic }}
\label{eq:axion-QED-hg-constraints}
\end{equation}
where $E_{\mathbb{Z}_L}$ is the energy scale at which operators charged under $\mathbb{Z}_L$ appear. The constraint $E_{\mathbb{Z}_L}\lesssim E_\text{winding}$ is satisfied by anomaly inflow. Namely, the inflow-induced charged zero modes on axion-strings allow one to construct local gauge-invariant operators that are charged under $\mathbb{Z}_L\subset SU(N_f)$, such as
\begin{equation}
    \mathcal{O}^I=\psi_\text{zero-mode}\Psi^I,\quad \widetilde{\mathcal{O}}_I=\bar\psi_\text{zero-mode}\widetilde{\Psi}_I
\end{equation}
where $\psi_\text{zero-mode}$ is an axion-string fermion zero mode and $I$ is the flavor index. Similarly, $E_{\mathbb{Z}_L}\lesssim E_\text{magnetic}$ is satisfied by anomaly inflow for monopoles. 

\subsection{Axion-QCD}

Next, let us discuss the case of axion Yang-Mills with charged matter (axion QCD), for which the action is given by Equation \ref{eq:axion-QED} with $F\wedge\star F\rightarrow 2\text{Tr}(F\wedge\star F)$ and $F\wedge F\rightarrow \text{Tr}(F\wedge F)$. As argued in \cite{Gaiotto2018,anomalies-coupling-constants}, while the 1-form center symmetry is no longer present, a mixed 't Hooft anomaly involving the $2\pi$-periodicity of $\theta$ is still present, due to the global structure of the symmetry group. In particular, for the case of $N_f$ fundamental quarks with equal mass, the full symmetry group that acts faithfully on the fermions is given by \footnote{If $(u,v,w)\in(SU(N),SU(N_f),U(1))$, then the subgroups $(\omega,1,\bar\omega)$, $(1,\omega_f,\bar\omega_f)$ and $(\omega,\omega_f,\overline{\omega\omega_f})$ for $\omega\in\mathbb{Z}_N$ and $\omega_f\in\mathbb{Z}_{N_f}$ all act trivially on the quark field.}
\begin{equation}
    \mathcal{G}=\frac{SU(N)\times SU(N_f)\times U(1)}{\mathbb{Z}_N\times\mathbb{Z}_{N_f}\times\mathbb{Z}_D}
\label{eq:axion-QCD-faithful}
\end{equation}
where $U(1)$ is the baryon number symmetry and $D=\text{lcm}(N,N_f)$. The non-trivial quotient in the global symmetry group $\mathcal{G}_\text{global}=\mathcal{G}/SU(N)$ allows one to activate 2-form background gauge fields which twist the $SU(N)$ and $SU(N_f)$ bundles into $PSU(N)$ and $PSU(N_f)$ bundles, respectively, and thereby enable fractional instanton number. \footnote{If $w_2$ and $w_2^{(f)}$ label the discrete magnetic fluxes of the dynamical and flavor groups, and $Z$ labels the fractional $U(1)$ flux, then the fluxes are related as $$
\int_{\Sigma_2} \frac{Z}{2 \pi}=\frac{1}{N} \int_{\Sigma_2} w_2+\frac{1}{N_f} \int_{\Sigma_2} w_2^{(f)}+n
$$where $n\in\mathbb{Z}$. The fractional instanton number is then $$
\frac{1}{8 \pi^2} \int \operatorname{Tr}(F \wedge F)=\frac{N(N-1)}{8\pi^2}\int B_c^{(2)}\wedge B_c^{(2)}=\frac{N(N-1)}{8 \pi^2}\int(Z-B_f^{(2)}) \wedge(Z-B_f^{(2)})
$$ where $B_c^{(2)}=2\pi w_2/N$ and $B_f^{(2)}=2\pi w_2^{(f)}/N_f$} Hence, below the scale at which operators charged under the $\mathbb{Z}_N$ global symmetry appear, the system is anomalous with respect to $2\pi$-shifts of the axion field, and the resulting higher-group extension of $U(1)_w^{(2)}$ suggests that
\begin{equation}
    E_{\mathbb{Z}_N} \lesssim E_\text{winding}
\end{equation}
where $E_{\mathbb{Z}_N}$ is the energy scale at which operators charged under $\mathbb{Z}_N\subset{U(N_f)}$ appear. Once again, this is consistent with anomaly inflow for axion strings. Namely, similar to the case of axion-QED, the charged zero modes of axion strings may be used to construct gauge-invariant operators that are charged under $\mathbb{Z}_N\subset{U(N_f)}$.

\section{Conclusions} \label{sec:conc}

In this work we have analyzed the phenomenological consequences of higher-form, higher-group, and non-invertible symmetries in axion effective field theories, with particular emphasis on the constraints imposed by symmetry emergence. Generally, the coupled background gauge transformations generated by 't Hooft anomalies and non-invertible fusion rules for ABJ anomalous symmetries imply parametric inequalities among the emergence scales of different generalized symmetries. These constraints are purely infrared statements: they follow from the structure of the effective theory and its anomalies, and do not depend on detailed assumptions about the ultraviolet completion.

We explored how these emergence constraints are realized in a variety of concrete settings in which axions are coupled to  Maxwell theory or various global structures of Yang–Mills, both with and without charged matter. Extending previous work on emergence in perturbative examples, we studied both 4d and 5d UV completions such as KSVZ-type models and higher-dimensional gauge theories, where the constraints are satisfied far from saturation due to the parametric separation of scales. More generally, we showed that anomaly inflow onto axion strings, monopoles, and their higher-dimensional analogues provides a universal and UV-independent mechanism that enforces all of the emergence constraints. In this sense, the same topological couplings that generate higher symmetries also guarantee their consistent emergence behavior.

From a phenomenological perspective, our results clarify how generalized symmetries constrain the range of validity of axion EFTs and the ordering of physical scales associated with strings, monopoles, charged matter, and axion shift breaking. These constraints organize a wide class of infrared effects—such as axion potentials generated by monopole loops or charged zero modes on defects—into a unified symmetry-based framework. More broadly, they provide concrete progress towards a generalized 't Hooft paradigm that organizes particle physics phenomena around their exact and approximate symmetries. 

Although we have focused on emergence phenomena in axion EFTs, the lessons learned here are more broadly applicable. In the 2+1d Goldstone-Maxwell model, vortices in the dynamical scalar fields once again carry charged zero modes \cite{Fukuda2021} and generate scalar potentials via loops, which is consistent with the higher-group and non-invertible symmetries identified in \cite{Damia2023}. For a generic finite group gauge theory with a topological action (Dijkgraaf-Witten twist), the general form of the gauged symmetry-protected topological phases attached to codimension-2 magnetic defects was derived in \cite{Barkeshli}. \footnote{The bulk topological phases in the magnetic brane couplings given in Equations \ref{eq:charges} and \ref{eq:charges-5d} are also examples of gauged SPTs. The gauged SPT attached to the magnetic monopole is reproduced in Equation 13 of \cite{Barkeshli}} Similar analysis was also applied to higher-form gauge theories in \cite{Hsin2022}. Finally, outside of the realm of topological terms and anomaly inflow, the standard model is rich with 't Hooft and ABJ anomalous flavor symmetries, and the resulting generalized symmetries may be associated with monopole-fermion scattering. In an upcoming work, we will make this connection explicit, and use this connection as a guide to exploring monopole-catalyzed baryon decay in grand unified theories. Ultimately, we anticipate that there are many more theories for which emergence constraints and the associated generalized symmetries may be fruitfully explored. 

\acknowledgments 

We would like to thank Prateek Agrawal, Daniel Brennan, Clay Cordova, Marius Kongsore, Seth Koren, Ho-Tat Lam, Ananth Malladi, Shu-Heng Shao, Ryan Thorngren, Chong Wang and Yifan Wang for useful discussions. We would also like to thank Marius Kongsore for helpful comments on the draft. The figures in this paper were generated using diagrams.net \cite{drawio}. This work was supported in part by the U.S.\ Department of Energy under the grant DE-SC0011702 and performed in part at the Kavli Institute for Theoretical Physics, supported by the National Science Foundation under Grant No.~NSF PHY-1748958.Part of this work was completed during the KITP program “Generalized Symmetries in Quantum Field Theory: High Energy Physics, Condensed Matter, and Quantum Gravity,” which is supported in part by Grant No. NSF PHY-2309135 to the KITP.

\appendix

\section{Mini-Review of Topological Order} \label{sec:TO}

Here, we review the defining properties and field theories of abelian topological orders. Such theories are essential in  constructing the non-invertible symmetry defects of axion-Maxwell, as discussed in Appendix \ref{sec:non-invertible}. 

\subsection{Fractional Quantum Hall State} \label{sec:FQH}

It is well-known in condensed matter theory that 2+1d topological orders are described by Chern-Simons theories. The defining physical property of topological order is the existence of fractionalized excitations (anyons). In particular, this refers to quasiparticles whose exchange yields a phase $e^{i\theta}$ for $\theta\neq0,\pi$ in their wavefunction. For a $\mathbb{Z}_N$ topological order with a single type of anyon, the appropriate Chern-Simons theory is 
\begin{equation}
    S_a=\frac{iN}{4\pi}\int_{\mathcal{M}_3} a^{(1)}\wedge da^{(1)}
\label{eq:pure-CS}
\end{equation}
where the $U(1)$ gauge field $a$ couples to the anyons. The $\theta=\pi/N$ self-statistics then follows from charge-flux attachment: the Chern-Simons term attaches $2\pi/N$ units of flux of the gauge field $a^{(1)}$ to a single anyon, as can be seen directly from the equations of motion. In particular, this implies that wrapping one anyon around another induces an Aharonov-Bohm phase of $e^{2\pi i/N}$.

A famous example of topological order is the fractional quantum Hall liquid. For an integer quantum Hall state with $k$ units of transverse Hall conductance, the appropriate action is a pure Chern-Simons theory for the electromagnetic gauge field $A^{(1)}$
\begin{equation}
    S_{\text{IQH}}=\frac{ik}{4\pi}\int_{\mathcal{M}_3} A^{(1)}\wedge dA^{(1)}
\label{eq:IQH}
\end{equation}
where quantization of $k$ follows from invariance under singular gauge transformations. Since $k$ must be quantized, we cannot simply take Equation \ref{eq:IQH} with $k=1/N$ as the action for a fractional quantum Hall state (Laughlin state with filling factor $\nu=1/N$). For this case, it turns out that the correct action is given by 
\begin{equation}
    S_\text{FQH}=i\int_{\mathcal{M}_3}\frac{N}{4\pi}a^{(1)}\wedge da^{(1)}+\frac{1}{2\pi}a^{(1)}\wedge dA^{(1)}
\label{eq:FQH}
\end{equation}
Naively substituting the equation of motion $A^{(1)}=-Na^{(1)}$ produces Equation \ref{eq:IQH} with $k=1/N$, and hence the desired fractional Hall conductivity. However, such a substitution is generally not valid, since the equation of motion is incompatible with the Dirac quantisation condition for $a^{(1)}$ \cite{tong2016lecturesquantumhalleffect}. Indeed, there is important information encoded in $S_\text{FQH}[a,A]$ that is not encoded in the naive action, including the presence of fractionalized quasiparticles, topological degeneracy, and the existence of chiral edge states \cite{Wen_QFT,tong2016lecturesquantumhalleffect}. The topological degeneracy can be seen directly from the algebra of the line operators for the anyons. For example, if we embed the state on a torus and define the line operators
\begin{equation}
    W_i=\exp\left (i\oint_{\gamma_i} a^{(1)}\right )
\label{eq:FQH-algebra}
\end{equation}
where $\gamma_1$ and $\gamma_2$ label two topologically distinct non-contractible strings on the torus, then it is straightforward to show that the line operators satisfy the algebra
\begin{equation}
    W_1 W_2=e^{2 \pi i / N} W_2 W_1
\end{equation}
following the equal-time commutation relations $\left[a_1(\mathrm{x}), a_2\left(\mathrm{x}^{\prime}\right)\right]=\frac{2 \pi i}{N} \delta^2\left(\mathrm{x}-\mathrm{x}^{\prime}\right)$. For a general $g$-genus surface, the ground state degeneracy is $N^g$.

Important for the discussion of non-invertible symmetries is the 't Hooft anomaly for the $\mathbb{Z}_N^{(1)}$ 1-form symmetry generated by $W$, which is implied by the algebra given in Equation \ref{eq:FQH-algebra}. To see this, let us take $B^{(2)}$ to be the background gauge field for the 1-form symmetry. Upon coupling Equation \ref{eq:pure-CS} to $B^{(2)}$, the action becomes
\begin{equation}
    S_{a}[B^{(2)}]=i\int_{\mathcal{M}_3}\frac{N}{4 \pi} a^{(1)} \wedge d a^{(1)}+\frac{N}{2 \pi} a^{(1)} \wedge B^{(2)}
\label{eq:minimal-TQFT-p=1}
\end{equation}
The 't Hooft anomaly then follows from the lack of invariance under the $\mathbb{Z}_N^{(1)}$ symmetry transformation $B^{(2)}\rightarrow B^{(2)}-d\Lambda^{(1)}$ and $a^{(1)}\rightarrow a^{(1)}+\Lambda^{(1)}$. The inflow action that is necessary to cancel this 't Hooft anomaly is given by the following 3+1d symmetry-protected topological (SPT) phase
\begin{equation}
    S_\text{inflow}=\frac{iN}{4\pi}\int_{\mathcal{M}_4}B^{(2)}\wedge B^{(2)}
\label{eq:FQH-inflow}
\end{equation}
where 
$\partial\mathcal{M}_4=\mathcal{M}_3$. Note that setting $B^{(2)}=dA^{(1)}/N$ in Equation \ref{eq:minimal-TQFT-p=1} reproduces Equation \ref{eq:FQH}. Then, the naive Lagrangian for the fractional quantum Hall state cancels the bulk anomaly inflow from Equation \ref{eq:FQH-inflow}. 

\subsection{$\mathbb{Z}_N$ Gauge Theory} \label{sec:ZNGT}

The above discussion has a natural generalization to mutual statistics in arbitrary spacetime dimension. For a $\mathbb{Z}_N$ topological order defined on a $D$-dimensional spacetime manifold $\mathcal{M}_D$, which describes $\mathbb{Z}_N$ statistics between point particles and $(D-2)$-dimensional extended objects, the appropriate topological term is the BF term
\begin{equation}
    S_\text{BF}=\frac{iN}{2\pi}\int_{\mathcal{M}_D} b^{(D-2)}\wedge da^{(1)},
\label{eq:BF}
\end{equation}
where $b^{(D-2)}$ couples to the extended objects and $a^{(1)}$ couples to point particles. Here, the relevant loop operators are the Wilson lines and 't Hooft membranes
\begin{equation}
    W(\Sigma_1)=\exp\left (i\int_{\Sigma_1}a^{(1)} \right),\quad V(\Sigma_{D-2})=\exp\left ( i\int_{\Sigma_{D-2}}b^{(D-2)} \right )
\end{equation}
for 1-dimensional and $(D-2)$-dimensional submanifolds $\Sigma_1$ and $\Sigma_{D-2}$ defined in space, respectively, and the associated $\mathbb{Z}_N$ symmetry algebra is given by
\begin{equation}
    W(\Sigma_1)V(\Sigma_{D-2})=e^{\frac{2\pi i}{N}\#\left ( \Sigma_1,\Sigma_{D-2}\right )}V(\Sigma_{D-2})W(\Sigma_1)
\end{equation}
where $\#(\Sigma_1,\Sigma_{D-2})$ is the intersection number of $\Sigma_1$ and $\Sigma_{D-2}$. This algebra implies a mixed 't Hooft anomaly between the $\mathbb{Z}_N^{(D-2)}$ symmetry generated by $W$ and the $\mathbb{Z}_N^{(1)}$ symmetry generated by $V$ \cite{brennan-hong,LUO2024}. To see this, let us take $B^{(D-1)}$ and $B^{(2)}$ to be background gauge fields for these $\mathbb{Z}_N^{(D-2)}$ and $\mathbb{Z}_N^{(1)}$ symmetries. Upon coupling the $\mathbb{Z}_N$ gauge theory to these background gauge fields, we have
\begin{equation}
    S_\text{BF}[B^{(D-1)},B^{(2)}]=\frac{iN}{2\pi}\int b^{(D-2)}\wedge da^{(1)}+a^{(1)}\wedge B^{(D-1)}+b^{(D-2)}\wedge B^{(2)}
\label{eq:BF-gauged}
\end{equation}
Under the $\mathbb{Z}_N^{(1)}$ symmetry transformation $B^{(2)}\rightarrow B^{(2)}-d\Lambda^{(1)}$ and $a^{(1)}\rightarrow a^{(1)}+\Lambda^{(1)}$, Equation \ref{eq:BF-gauged} is not invariant due to the second term. Similarly, under a $\mathbb{Z}_N^{(D-2)}$ symmetry transformation, Equation \ref{eq:BF-gauged} lacks invariance due to the third term. The inflow action that is necessary to cancel this mixed 't Hooft anomaly is given by the following bulk SPT phase
\begin{equation}
    \frac{iN}{2\pi}\int_{\mathcal{M}_{D+1}} B^{(D-1)}\wedge B^{(2)}
\end{equation}
where $\partial\mathcal{M}_{D+1}=\mathcal{M}_D$.

\section{Discrete symmetries of axion-Maxwell} \label{sec:non-invertible}

Here, we review the non-invertible axion-shift and electric 1-form symmetries for 3+1d and 4+1d axion-Maxwell. For general $\alpha$, the invertible symmetries generated by the charges in Equations \ref{eq:charges} and \ref{eq:charges-5d} and reviewed in Appendix \ref{sec:invertible} lack gauge invariance. For $\alpha\in 2\pi \mathbb{Q}/\mathbb{Z}$, one can write modified gauge-invariant symmetry operators, which naively reduce to the original invertible symmetry operators upon integrating out the new auxiliary fields. Such operators are generally non-invertible, and their topological nature can be formally justified through half-$p$-form gauging constructions \cite{SHS-non-invertible,SHS-non-invertible-II,Roumpedakis2023}, which we review below. Both the non-invertible fusion algebras and the half-$p$-form gauging constructions may be used to arrive to the emergence constraints summarized in Appendix \ref{sec:emergence} and stated in the main text.

\subsection{Invertible symmetries} \label{sec:invertible}

\subsubsection{3+1d axion-Maxwell}

For strictly invertible symmetries, the appropriate statement is that on a spacetime with non-trivial topology, the $U(1)^{(0)}$ axion-shift symmetry is broken down to $\mathbb{Z}_K^{(0)}$. At the level of the Lagrangian, this follows from the fact that the topological term given in Equation \ref{eq:axion-Maxwell-3+1d} is only invariant under discrete shifts of the axion field $\delta({a}/{f})={2 \pi n}/{K}$ due to the instanton number
\begin{equation}
    \frac{1}{8 \pi^2} \int F \wedge F \in \mathbb{Z}
\end{equation}
Alternatively, we can also argue for the $\mathbb{Z}_K^{(0)}$ axion-shift symmetry at the level of the symmetry defect. Namely, the symmetry defect that implements the axion-shift symmetry is of the form
\begin{equation}
    \mathcal{U}_\alpha\left(\Sigma_3\right)=\exp \left[i \alpha\left(\int_{\Sigma_3} \star J^{(1)}-\frac{K}{8 \pi^2} A \wedge F\right)\right]
\label{eq:axion-shift-defect}
\end{equation}
where the Chern-Simons term ensures that the modified current is closed and hence that $U_\alpha\left(\Sigma_3\right)$ is topological. Here, we must have that $\alpha=2 \pi n / K$ for $n \in \mathbb{Z}$ for the coefficient on the Chern-Simons term to be properly quantized. Note that such quantization is required for invariance of the Chern-Simons term under singular gauge transformations, which are only possible in the presence of non-trivial cycles. In our previous argument, the relevance of non-trivial spacetime topology in fixing $\alpha \in \mathbb{Z}_K$ was in the role of abelian instantons. Namely, we have that $\pi_3(U(1))=0$. As noted in \cite{Yokokura}, the role of abelian instantons in enforcing $\alpha\in\mathbb{Z}_K$ at the level of the symmetry defect becomes apparent by noting that the Chern-Simons term must be independent of its bulk extension
\begin{equation}
    \exp \left[\frac{i K \alpha}{8 \pi^2} \int_{\Sigma_4} F \wedge F\right] =\exp \left[\frac{i K \alpha}{8 \pi^2} \int_{\Sigma_4^{\prime}} F \wedge F\right]\rightarrow\exp \left[\frac{i K \alpha}{8 \pi^2} \int_{\Sigma} F \wedge F\right]=1
\end{equation}
where $\partial\Sigma_4=\partial\Sigma_4'=\Sigma_3$ and $\Sigma=\Sigma_4\cup\overline{\Sigma}_4'$. 

The electric 1-form symmetry for the abelian gauge field $A$ is also broken down to $\mathbb{Z}_K$. In 3+1d, the appropriate symmetry defect is
\begin{equation}
    \mathcal{U}_\alpha\left(\Sigma_2\right)=\exp \left[i \alpha\left(\int_{\Sigma_2} \star J_e^{(2)}+\frac{K}{4 \pi^2 f} a F\right)\right]
\label{eq:X-invertible}
\end{equation}
where we again have that $\alpha=2 \pi n / K$ for $n \in \mathbb{Z}$. Here, the fact that $n$ is integer-valued ensures that $U_\alpha\left(\Sigma_2\right)$ is well-defined under axion shifts $a \rightarrow a+2 \pi f$, since $\int_{\Sigma_2} F \in 2 \pi \mathbb{Z}$.

\subsubsection{4+1d axion-Maxwell}

In 4+1d, this $\mathbb{Z}_K$ axion-shift symmetry becomes a $\mathbb{Z}_K$ electric 1-form symmetry for the abelian gauge field $C$ whose zero mode is the axion. The corresponding invertible symmetry defect takes the form of Equation \ref{eq:axion-shift-defect} with $J^{(1)}\rightarrow J_e^{(2)}$.

On the other hand, the appropriate symmetry defect for the $\mathbb{Z}_K$ electric 1-form symmetry for $A$ is
\begin{equation}
    \mathcal{U}_\alpha\left(\Sigma_3\right)=\exp \left[i \alpha\left(\int_{\Sigma_3} \star J_e^{(2)}+\frac{K}{4 \pi^2} C \wedge F\right)\right]
\end{equation}
As above, $\alpha\in\mathbb{Z}_K$ follows from requiring that the coefficient on the Chern-Simons term be properly quantized.

\subsection{Non-invertible symmetries} \label{sec:non-invertible}

\subsubsection{3+1d axion-shift symmetry} \label{sec:axion-shift-non-invertible}

As demonstrated in \cite{SHS-non-invertible,SHS-non-invertible-II,Cordova-non-invertible}, it is possible to construct a non-invertible axion shift symmetry that holds for any $\alpha=2 \pi p / K N$ where $p / N \in \mathbb{Q} / \mathbb{Z}$. To motivate such symmetry defects, let us consider the case of $K=p=1$, for which the Chern-Simons term in Equation \ref{eq:axion-shift-defect} is
\begin{equation}
    -\frac{i}{4 \pi N} \oint_{\Sigma_3} A \wedge F
\label{eq:naive-cs}
\end{equation}
As discussed in Appendix \ref{sec:FQH}, this is a natural first attempt at writing down the field theory for a fractional quantum Hall (Laughlin) state. While Equation \ref{eq:naive-cs} produces the correct Hall conductance, it lacks invariance under singular gauge transformations. Instead, the correct gauge-invariant action for the fractional quantum Hall state is
\begin{equation}
    i \oint_{\Sigma_3}\left(\frac{N}{4 \pi} c \wedge d c+\frac{1}{2 \pi} c \wedge F\right)
\end{equation}
where $c$ is a dynamical $U(1)$ gauge field that couples to fractionalized quasiparticles of the theory. This motivates the following symmetry defect for a $p=1$ transformation
\begin{equation}
    \mathcal{D}_{p=1}\left(\Sigma_3\right)=\left.\int Dc \right|_{\Sigma_3} \exp \left[i \oint_{\Sigma_3}\left(\frac{2 \pi}{N} \star J^{(1)}+\frac{N}{4 \pi} c \wedge d c+\frac{1}{2 \pi} c \wedge F\right)\right]
\label{eq:fqh-defect}
\end{equation}
A formal justification of the topological nature of this non-invertible symmetry defect follows from the half-gauging procedure presented in \cite{SHS-non-invertible,Cordova-non-invertible}. Let us review this here, since this will be useful for understanding the 4+1d picture.

Let us first consider performing an axion-shift $a\rightarrow a-\frac{2\pi p f}{N}$ on the entire spacetime manifold $\mathcal{M}_4$. By the ABJ anomaly for the axion-shift symmetry, such a rotation generates an anomalous phase of the form $\frac{ip}{4\pi N}\int_{\mathcal{M}_4}F\wedge F$. To cancel this phase, one may gauge the discrete $\mathbb{Z}_N^{(1)}$ subgroup of the magnetic 1-form symmetry $U(1)_m^{(1)}$ with the following 2-form gauge theory
\begin{equation}
    \delta S_\text{gauging}=\int_{\Sigma_4}\frac{i}{2 \pi} b \wedge F+\frac{iN}{2 \pi} b \wedge d c+\frac{iN p'}{4 \pi} b \wedge b
\end{equation}
where $p p^{\prime}=1 \bmod N$. Here, the first term is the minimal coupling term for the magnetic symmetry
\begin{equation}
    i b \wedge \star J_m^{(2)}=\frac{i}{2 \pi} b \wedge F 
\end{equation}
Furthermore, the second term enforces $db=0$ and restricts the holonomy of $b$ to be $\mathbb{Z}_N$-valued, hence making $b$ a $\mathbb{Z}_N$ gauge field. Finally, the third term describes a one-form SPT \cite{Gaiotto2015,Thorngren2015,Kapustin2017} and is added for the purpose explained above. Namely, upon integrating out $b$ and $c$, one finds
\begin{equation}
    \delta S_\text{gauging}=-\frac{ip}{4\pi N}\int_{\mathcal{M}_4}F\wedge F
\end{equation}
which precisely cancels the anomalous phase generated by $a\rightarrow a-\frac{2\pi p f}{N}$. Hence, we see that the combined operation of performing an axion-shift transformation and gauging the appropriate discrete subgroup of the magnetic 1-form symmetry is a symmetry of the theory. 

We can construct the topological defect that implements this symmetry by performing the same operations on half of spacetime \cite{D1,D2,D3}. In particular, we restrict the axion-shift transformation to a 4-dimensional submanifold $\Sigma_4$, such that $a\rightarrow a-\frac{2\pi p f}{N}\delta^{(0)}(\Sigma_4)$, and gauge $\mathbb{Z}_N^{(1)}\subset U(1)_m^{(1)}$ on $\Sigma_4$. From the minimal coupling for the axion-shift current, the axion-shift transformation now generates an additional contribution of the form $\oint_{\Sigma_3}\frac{2\pi i p}{N}\star J^{(1)}$ where $\Sigma_3=\partial\Sigma_4$ \footnote{For an $n$-dimensional submanifold $\Sigma_n$ of a $D$-dimensional manifold $\mathcal{M}^D$, the $(D-n)$-form $\delta^{(D-n)}(\Sigma_n)$ has the defining property that $$
\int_{\mathcal{M}^D} \delta^{(D-n)}(\Sigma_n) \wedge A^{(n)}=\int_{\Sigma_n} A^{(n)}
$$The contribution $\oint_{\Sigma_3}\frac{2\pi i p}{N}\star J^{(1)}$ then follows from $$
\delta^{(D-n+1)}(\partial \Sigma_n)=(-1)^{(D-n+1)} d \delta^{(D-n)}(\Sigma_n)
$$which follows from Stokes' theorem.}. Furthermore, the remaining terms no longer exactly cancel, but instead yield a minimal $\mathbb{Z}_N$ TQFT on the boundary $\Sigma_3$. To be precise, as demonstrated in \cite{SHS-non-invertible}, we have that
\begin{equation}
    \int_{\Sigma_4}\left(\frac{i p}{4 \pi N} F \wedge F+\frac{i}{2 \pi} b \wedge F+\frac{i N}{2 \pi} b \wedge d c+\frac{i N k}{4 \pi} b \wedge b\right)=\oint_{\Sigma_3} \mathcal{A}^{N, p}[d A / N] .
\label{eq:minimal-TQFT}
\end{equation}
The minimal $\mathbb{Z}_N$ TQFT $\mathcal{A}^{N,p}$ is the natural generalization of the TQFT for the Laughlin state \cite{Gaiotto2015,Seiberg2019}. It is characterized by symmetry lines $W^s$ of topological spin \footnote{For a spin QFT, the topological spin of the symmetry line $W^s$ is $h\left[W^s\right]=\frac{p s^2}{2 N}\bmod\frac{1}{2}$, since we can always dress the operator with a transparent spin-half line.}
\begin{equation}
    h\left[W^s\right]=\frac{p s^2}{2 N} \bmod 1
\end{equation}
where $p$ and $N$ are coprime and $W^N=1$. The coprime condition ensures that only the trivial line $1$ braids trivially with all other lines, hence the name \textit{minimal} TQFT. Such theories possess a 't Hooft anomaly characterized by the inflow action
\begin{equation}
    S_{\text {inflow }}=-\frac{i p N}{4\pi} \int_{\Sigma_4} B^{(2)}\wedge B^{(2)}
\end{equation}
where $B^{(2)}$ is a two-form background gauge field for the $\mathbb{Z}_N^{(1)}$ 1-form symmetry generated by the Wilson line $W$. Ultimately, we find that the resulting symmetry defect takes the form
\begin{equation}
    \mathcal{D}_{p/N}(\Sigma_3)=\exp \left[\oint_{\Sigma_3}\left(\frac{2 \pi i p}{N} \star J^{(1)}+\mathcal{A}^{N, p}[d A / N]\right)\right] ,
\label{eq:3+1d-axion-shift-non-invertible}
\end{equation}
which satisfies a non-invertible fusion algebra. For the case of $p=1$, the minimal TQFT $\mathcal{A}^{N,1}$ takes the familiar form of the fractional quantum Hall state, and the corresponding symmetry defect takes the form of Equation \ref{eq:fqh-defect}. In this case, the fusion algebra is given by
\begin{equation}
\mathcal{D}_{{1}/{N}}(\Sigma_3) \times \mathcal{D}_{{1}/{N}}^{\dagger}(\Sigma_3)=\exp \left[i \oint_M\left(\frac{N}{4 \pi} c \wedge d c-\frac{N}{4 \pi} \bar{c} \wedge d \bar{c}+\frac{1}{2 \pi}(c-\bar{c}) \wedge F\right)\right]
\end{equation}
where $\mathcal{D}_{\alpha}^\dagger\equiv\mathcal{D}_{-\alpha}$ and the right-hand side is a condensation operator from 1-gauging $\mathbb{Z}_N^{(1)}\subset U(1)_m^{(1)}$ (see \cite{Roumpedakis2023,C1,C2,C3,C4,C5} for references on condensation operators). Finally, note that the smoothness condition imposed by the BF term in Equation \ref{eq:minimal-TQFT} ensures that the imposed Dirichlet boundary condition for $b$ is topological and hence that the non-invertible defect in Equation \ref{eq:3+1d-axion-shift-non-invertible} is topological \cite{SHS-non-invertible}.

\subsubsection{4+1d electric 1-form symmetry for $C$} \label{sec:4+1d-C-non-invertible}

\begin{figure}
    \centering
    \includegraphics{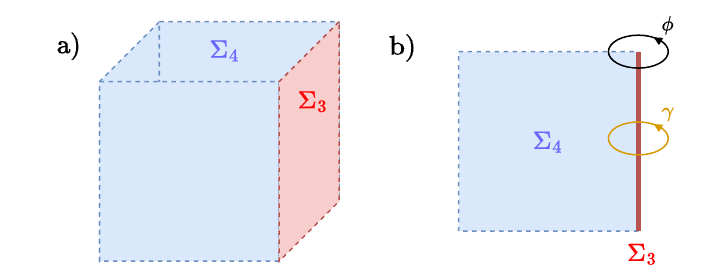}
    \caption{a) 0-form gauging of magnetic symmetry in 3+1d. b) 1-form gauging of magnetic symmetry in 4+1d. Any Wilson line $L(\gamma)$ that has non-trivial linking with $\Sigma_3$ (or non-trivial intersection number with $\Sigma_4$) will transform non-trivially under the insertion of $\mathcal{D}_{p/N}(\Sigma_3)$ given in Equation \ref{eq:4+1d-C-non-invertible}.}
    \label{fig:higher-gauging}
\end{figure}

In 4+1d, the non-invertible axion-shift symmetry becomes a non-invertible electric 1-form symmetry for $C$. Following \cite{Aguilera}, the symmetry defect for this electric 1-form symmetry can be constructed in a manner analogous to above. Namely, to construct a non-invertible symmetry defect for the electric 1-form symmetry for $C$ on a 3-manifold $\Sigma_3$, we perform an electric 1-form symmetry transformation $C\rightarrow C-\frac{2\pi p }{N}\delta^{(1)}(\Sigma_4)$ where $\partial\Sigma_4=\Sigma_3$ \footnote{Under such a transformation, any Wilson line $L(\gamma)$ for which $\#\left(\gamma, \Sigma_4\right)=\operatorname{Link}\left(\gamma, \Sigma_3\right) \neq 0$ will transform non-trivially. Note that this electric 1-form symmetry transformation is equivalent to a transformation of $C$ by a 0-form gauge parameter with non-trivial holonomy around $\Sigma_3$: $C\rightarrow C+d\Lambda^{(0)}$ where $d^2\Lambda^{(0)}=\frac{2\pi p}{N}\delta^{(2)}(\Sigma_3)$ or $\Lambda^{(0)}(\phi+2 \pi)=\Lambda^{(0)}(\phi)+\frac{2 \pi p}{N}$ if $\Lambda^{(0)}$ is locally parameterized by $\phi$ along $\gamma$. See Figure \ref{fig:higher-gauging}.}, and gauge $\mathbb{Z}_N^{(2)}\subset U(1)_m^{(2)}$ on $\Sigma_4$. Since $\Sigma_4$ is a codimension-1 object, this is an example of half-higher-gauging \cite{Roumpedakis2023} (see Figure \ref{fig:higher-gauging}). The resulting defect is
\begin{equation}
    \mathcal{D}_{p/N}(\Sigma_3)=\exp \left[\oint_{\Sigma_3}\left(\frac{2 \pi i p}{N} \star J_e^{(2)}+\mathcal{A}^{N, p}[d A / N]\right)\right] ,
\label{eq:4+1d-C-non-invertible}
\end{equation}

\subsubsection{3+1d electric 1-form symmetry for $A$} \label{sec:3+1d-X-non-invertible}

An analogous discussion holds for the electric 1-form symmetry for $A$. The symmetry defect presented in Equation \ref{eq:X-invertible} is not valid for any $\alpha=2\pi p/KN$ since it is not invariant under $2\pi$ shifts of the axion field. The appropriate shift-invariant modification (for $K=1$) is \footnote{Invariance under $2\pi$-shifts of $a$ and $\chi$ follows from $\int_{\Sigma_2}F\in2\pi\mathbb{Z}$ and $\int_{\Sigma_2}du\in2\pi\mathbb{Z}$.} \cite{SHS-non-invertible-II,Yokokura}
\begin{equation}
    \mathcal{D}_{{p}/{N}}^{(1)}\left(\Sigma_2\right)=\left.\int D \chi D c\right|_{\Sigma_2} \exp \left[\oint_{\Sigma_2}\left(\frac{2 \pi i p}{N} \star J_e^{(2)}+\frac{i N}{2 \pi} \chi d c+\frac{i p}{2 \pi f} a d c+\frac{i}{2 \pi} \chi F \right)\right]
\label{eq:X-defect}
\end{equation}
where $\chi$ is a dynamical compact scalar field and $c$ is a dynamical $U(1)$ gauge field. This is again motivated by the observation that naively integrating out $c$ and $\chi$ yields the symmetry defect given in Equation \ref{eq:X-invertible} for $\alpha=2\pi p/N$. Similar to the case of the axion-shift symmetry, one can formally show that Equation \ref{eq:X-defect} is topological following the half-higher-gauging construction presented in \cite{SHS-non-invertible-II}. The final three terms in Equation \ref{eq:X-defect} describe a 1+1d $\mathbb{Z}_N$ gauge theory, as described in Appendix \ref{sec:ZNGT}, which possesses a 't Hooft anomaly characterized by inflow action
\begin{equation}
    S_{\text {inflow }}=-\frac{i N}{2 \pi} \int_{\Sigma_3} B^{(1)}  B^{(2)}
\end{equation}
where $B^{(1)}$ and $B^{(2)}$ are background gauge fields for $\mathbb{Z}_N^{(0)}$ and $\mathbb{Z}_N^{(1)}$ symmetries. Here, the $\mathbb{Z}_N^{(0)} \times \mathbb{Z}_N^{(1)}$ global symmetry is generated by the operators $W(\gamma)=e^{i \oint_\gamma c}$ and $V(x)=e^{i \chi(x)}$ and the 't Hooft anomaly is implied by the clock-and-shift algebra $W V=e^{2 \pi i / N} V W$.

\subsubsection{4+1d electric 1-form symmetry for $A$} \label{sec:4+1d-non-invertible-X}

To construct the symmetry defect for the electric 1-form symmetry of $A$ for the 4+1d case, we follow the appropriate analogue of the half-higher-gauging construction for the 3+1d electric 1-form symmetry in \cite{SHS-non-invertible-II}. For the 3+1d case, it was shown that construction of such a defect requiring 1-gauging the $\mathbb{Z}_N^{(1)}\times \mathbb{Z}_N^{(2)}$ subgroup of the $U(1)_m^{(1)}\times U(1)_w^{(2)}$ symmetry on a codimension-1 manifold $\Sigma_3$ with the appropriate choice of discrete torsion. In 4+1d, both the magnetic 1-form symmetry and 2-form winding symmetry become magnetic 2-form symmetries. Hence, in this case, we gauge $\mathbb{Z}_N^{(2)}\times\mathbb{Z}_N^{(2)}\subset U(1)_m^{(2)}\times U(1)_m^{(2)}$ on a codimension-1 manifold $\Sigma_4$. The appropriate condensation defect is
\begin{equation}
    \mathcal{C}\left(\Sigma_4\right)=\frac{\left|H_{\partial}^0\right|^2}{\left|H_{\partial}^1\right|^2} \sum_{b_A^{(2)}, b_C^{(2)} \in H_{\partial}^2} \exp \left[\frac{2 \pi i}{N} \int_{\Sigma_4}\left(b_A^{(2)} \cup b_C^{(2)}+b_A^{(2)} \cup \frac{F_A}{2 \pi}-p b_C^{(2)} \cup \frac{F_C}{2 \pi}\right)\right]
\end{equation}
where $H_\partial^i=H^i(\Sigma_4,\partial\Sigma_4;\mathbb{Z}_N)$ labels topologically distinct classes of gauge fields that vanish on $\partial\Sigma_4$ and we have now defined $F_A=dA$ and $F_C=dC$. We may rewrite this as 
\begin{align}
\mathcal{C}\left(\Sigma_{4}\right)&= \exp \left(\frac{2 \pi i p}{N} \int_{\Sigma_{4}} \frac{F_C}{2 \pi} \wedge \frac{F_A}{2 \pi}\right) \times \\
& \frac{\left|H_{\partial}^0\right|^2}{\left|H_{\partial}^1\right|^2} \sum_{b_A^{(2)},b_C^{(2)} \in H_{\partial}^2} \exp \left[\frac{2 \pi i}{N} \int_{\Sigma_{4}}\left(b_A^{(2)}-p \frac{F_C}{2 \pi}\right) \cup\left(b_C^{(2)}+\frac{F_A}{2 \pi}\right)\right] .
\end{align}
Following the non-conservation equation for the electric 1-form symmetry current for $A$, the first line becomes
\begin{equation}
    \exp \left(\frac{2 \pi i p}{N} \int_{\Sigma_{4}} \frac{F_C}{2 \pi} \wedge \frac{F_A}{2 \pi}\right)=\exp \left(\frac{2 \pi i p}{N} \oint_{\Sigma_{3}} \star J_{e}^{(2)}\right)
\end{equation}
where $\partial\Sigma_4=\Sigma_3$. Furthermore, following the natural generalization of the discussion in Appendix B of \cite{SHS-non-invertible-II}, one can check that
\begin{align}
&\frac{\left|H_{\partial}^0\right|^2}{\left|H_{\partial}^1\right|^2} \sum_{b_A^{(2)},b_C^{(2)} \in H_{\partial}^2} \exp \left[\frac{2 \pi i}{N} \int_{\Sigma_{4}}\left(b_A^{(2)}-p \frac{F_C}{2 \pi}\right) \cup\left(b_C^{(2)}+\frac{F_A}{2 \pi}\right)\right]\\
&=\int DcD\tilde{c}|_{\Sigma_3}\exp\left [ \oint_{\Sigma_3}\left ( \frac{i N}{2 \pi} \tilde{c}\wedge d c+\frac{i p}{2 \pi} C\wedge d c+\frac{i}{2 \pi} c\wedge F \right ) \right ]
\end{align}
where $c$ and $\tilde{c}$ are dynamical $U(1)$ gauge fields. The right hand side is a 2+1d $\mathbb{Z}_N$ gauge theory, which possesses a 't Hooft anomaly characterized by inflow action
\begin{equation}
    S_\text{inflow}=-\frac{i N}{2 \pi} \int_{\Sigma_{3}} \tilde{B}^{(2)}\wedge B^{(2)}
\end{equation}
where $\tilde{B}^{(2)}$ and $B^{(2)}$ are background gauge fields for the $\mathbb{Z}_N^{(1)}$ symmetries. Here, the $\mathbb{Z}_N^{(1)} \times \mathbb{Z}_N^{(1)}$ global symmetry is generated by the operators $W(\gamma)=e^{i \oint_\gamma c}$ and ${V}(\gamma)=e^{i\oint_\gamma \tilde{c}}$ and the 't Hooft anomaly is implied by the clock-and-shift algebra $W(\gamma)V(\gamma')={e^{2\pi i/N}}V(\gamma')W(\gamma)$ for $\#(\gamma,\gamma')=1\bmod 2$.

Hence, the appropriate gauging of the magnetic 2-form symmetries yields the desired non-invertible symmetry defect for the electric 1-form symmetry. Namely, we have that $\mathcal{C}\left(\Sigma_{4}, \partial \Sigma_{4}=\Sigma_{3}\right)=\mathcal{D}_{p / N}^{(1)}\left(\Sigma_{3}\right)$.

\subsection{Emergence conjecture for non-invertible symmetries} \label{sec:emergence}

\begin{table}[!htbp]
\begin{tabular}{|l|l|l|l|}
\hline
                  & Non-invertible symmetry & Symmetry defect & Gauged symmetries \\ \hline
\multirow{2}{*}{} 3+1d & Axion-shift & $\mathcal{D}_{p/N}\left(\Sigma_3\right)=U_{p/N} \times \mathcal{A}^{N, p}(A)$ & $U(1)_m^{(1)}$ \\ \cline{2-4} 
                  & Electric 1-form for $A$ & $\mathcal{D}_{p/N}\left(\Sigma_2\right)=U_{p/N} \times \mathcal{Z}_{1}^{N, p}(A,a)$ & $U(1)_m^{(1)}\times U(1)_w^{(2)}$ \\ \hline
\multirow{2}{*}{} 4+1d & Electric 1-form for $C$ & $\mathcal{D}_{p/N}\left(\Sigma_3\right)=U_{p/N} \times \mathcal{A}^{N, p}(A)$ & $U(1)_m^{(2)}$ \\ \cline{2-4} 
                  & Electric 1-form for $A$ & $\mathcal{D}_{p/N}\left(\Sigma_3\right)=U_{p/N} \times \mathcal{Z}_{2}^{N, p}(A,C)$ & $U(1)_m^{(2)}\times U(1)_m^{(2)}$ \\ \hline
\end{tabular}
\caption{Summary of non-invertible symmetry defects for 3+1d and 4+1d axion-Maxwell. $U_{p/N}$ is the naive invertible symmetry defect defined in terms of the relevant anomalous current. $\mathcal{A}^{N,p}$ is the minimal $\mathbb{Z}_N$ TQFT defined in Appendix \ref{sec:axion-shift-non-invertible} while $\mathcal{Z}_d^{N,p}$ is the $d$-spacetime dimensional $\mathbb{Z}_N$ gauge theory described in Appendix \ref{sec:ZNGT} and \ref{sec:4+1d-non-invertible-X}. The rightmost column refers to the symmetries that must be gauged in the construction of these non-invertible defects. }
\label{table:non-invertible}
\end{table}

The non-invertible symmetries reviewed in this appendix are summarized in Table \ref{table:non-invertible}. The resulting emergence constraints for 3+1d axion-Maxwell are given by
\begin{equation}
    E_{\text {electric }} \lesssim \min \left\{E_{\text {winding }}, E_{\text {magnetic }}\right\} \quad, \quad E_{\text {shift }} \lesssim E_{\text {magnetic }} .
\end{equation}
Given the discussion above, one can argue for the former based on the observation that the construction of the non-invertible symmetry defect for the electric 1-form symmetry requires half-higher-gauging the magnetic 1-form and 2-form winding symmetries. Similarly, the latter follows from the observation that the construction of the non-invertible symmetry defect for the axion-shift symmetry requires half-gauging the magnetic 1-form symmetry.

Alternatively, one can argue for these constraints using the fusion rules for these defects. Namely, it is shown in \cite{SHS-non-invertible-II} that the fusion of two non-invertible axion-shift symmetry defects defined on a 3-manifold $\Sigma_3$ yields a sum over magnetic 1-form symmetry defects defined on closed 2-cycles of $\Sigma_3$. Similarly, it is shown that the fusion of two non-invertible electric 1-form symmetry defects yields sums over 2-form winding and magnetic 1-form symmetry defects. 

Following similar arguments, the emergence constraints for 4+1d axion-Maxwell are given by
\begin{equation}
    E_{\text{electric}}^{(A)} \lesssim \min\{E_{\text{magnetic}}^{(C)},E_{\text{magnetic}}^{(A)}\},\quad E_{\text{electric}}^{(C)} \lesssim E_{\text{magnetic}}^{(A)}
\end{equation}
The former follows from the observation that construction of the non-invertible symmetry defect for the electric 1-form symmetry for $A$ requires half-higher-gauging of the magnetic 2-form symmetries for $C$ and $A$, or from the observation that the fusion of such defects yields sums over magnetic 2-form symmetry defects. Similarly, the latter follows from the observation that construction of the non-invertible symmetry defect for the electric 1-form symmetry for $C$ requires half-higher-gauging of the magnetic 2-form symmetry for $A$, or from the observation that the fusion of such defects defined on a $3$-manifold $\Sigma_3$ yields a sum over magnetic 2-form symmetry defects defined on closed 2-cycles of $\Sigma_3$.

 \bibliographystyle{JHEP}
 \bibliography{biblio.bib}

@Article{emergent-hg-brennan,
author={Brennan, T. Daniel
and C{\'o}rdova, Clay},
title={Axions, higher-groups, and emergent symmetry},
journal={Journal of High Energy Physics},
year={2022},
month={Feb},
day={17},
volume={2022},
number={2},
pages={145},
issn={1029-8479},
doi={10.1007/JHEP02(2022)145},
url={https://doi.org/10.1007/JHEP02(2022)145}
}

@article{brennan-hong,
    author = "Brennan, T. Daniel and Hong, Sungwoo",
    title = "{Introduction to Generalized Global Symmetries in QFT and Particle Physics}",
    eprint = "2306.00912",
    archivePrefix = "arXiv",
    primaryClass = "hep-ph",
    month = "6",
    year = "2023"
}

@article{bhardwaj2023lecturesgeneralizedsymmetries,
    author = "Bhardwaj, Lakshya and Bottini, Lea E. and Fraser-Taliente, Ludovic and Gladden, Liam and Gould, Dewi S. W. and Platschorre, Arthur and Tillim, Hannah",
    title = "{Lectures on generalized symmetries}",
    eprint = "2307.07547",
    archivePrefix = "arXiv",
    primaryClass = "hep-th",
    doi = "10.1016/j.physrep.2023.11.002",
    journal = "Phys. Rept.",
    volume = "1051",
    pages = "1--87",
    year = "2024"
}

@article{McGreevy,
   author = "McGreevy, John",
   title = "Generalized Symmetries in Condensed Matter", 
   journal= "Annual Review of Condensed Matter Physics",
   year = "2023",
   volume = "14",
   number = "Volume 14, 2023",
   pages = "57-82",
   doi = "https://doi.org/10.1146/annurev-conmatphys-040721-021029",
   url = "https://www.annualreviews.org/content/journals/10.1146/annurev-conmatphys-040721-021029",
   publisher = "Annual Reviews",
   issn = "1947-5462",
   type = "Journal Article",
  }

@article{Schafernameki2023,
    author = "Schafer-Nameki, Sakura",
    title = "{ICTP lectures on (non-)invertible generalized symmetries}",
    eprint = "2305.18296",
    archivePrefix = "arXiv",
    primaryClass = "hep-th",
    doi = "10.1016/j.physrep.2024.01.007",
    journal = "Phys. Rept.",
    volume = "1063",
    pages = "1--55",
    year = "2024"
}

@inproceedings{Cordova2022,
    author = "Cordova, Clay and Dumitrescu, Thomas T. and Intriligator, Kenneth and Shao, Shu-Heng",
    title = "{Snowmass White Paper: Generalized Symmetries in Quantum Field Theory and Beyond}",
    booktitle = "{Snowmass 2021}",
    eprint = "2205.09545",
    archivePrefix = "arXiv",
    primaryClass = "hep-th",
    month = "5",
    year = "2022"
}

@misc{Baez2004,
      title={Higher Gauge Theory: 2-Connections on 2-Bundles}, 
      author={John Baez and Urs Schreiber},
      year={2004},
      eprint={hep-th/0412325},
      archivePrefix={arXiv},
      primaryClass={hep-th},
      url={https://arxiv.org/abs/hep-th/0412325}, 
}

@Article{Baez2011,
author={Baez, John C.
and Huerta, John},
title={An invitation to higher gauge theory},
journal={General Relativity and Gravitation},
year={2011},
month={Sep},
day={01},
volume={43},
number={9},
pages={2335-2392},
issn={1572-9532},
doi={10.1007/s10714-010-1070-9},
url={https://doi.org/10.1007/s10714-010-1070-9}
}

@misc{Gukov2013,
      title={Topological Quantum Field Theory, Nonlocal Operators, and Gapped Phases of Gauge Theories}, 
      author={Sergei Gukov and Anton Kapustin},
      year={2013},
      eprint={1307.4793},
      archivePrefix={arXiv},
      primaryClass={hep-th},
      url={https://arxiv.org/abs/1307.4793}, 
}

@Article{Bhardwaj2024-II,
	title={{Generalized charges, part I: Invertible symmetries and higher representations}},
	author={Lakshya Bhardwaj and Sakura Schäfer-Nameki},
	journal={SciPost Phys.},
	volume={16},
	pages={093},
	year={2024},
	publisher={SciPost},
	doi={10.21468/SciPostPhys.16.4.093},
	url={https://scipost.org/10.21468/SciPostPhys.16.4.093},
}

@Article{Bartsch2024-II,
	title={{Non-invertible symmetries and higher representation theory II}},
	author={Thomas Bartsch and Mathew Bullimore and Andrea E. V. Ferrari and Jamie Pearson},
	journal={SciPost Phys.},
	volume={17},
	pages={067},
	year={2024},
	publisher={SciPost},
	doi={10.21468/SciPostPhys.17.2.067},
	url={https://scipost.org/10.21468/SciPostPhys.17.2.067},
}

@article{Bartsch2023,
    author = "Bartsch, Thomas and Bullimore, Mathew and Grigoletto, Andrea",
    title = "{Higher representations for extended operators}",
    eprint = "2304.03789",
    archivePrefix = "arXiv",
    primaryClass = "hep-th",
    month = "4",
    year = "2023"
}

@Article{Bhardwaj2023,
	title={{Non-invertible higher-categorical symmetries}},
	author={Lakshya Bhardwaj and Lea E. Bottini and Sakura Schäfer-Nameki and Apoorv Tiwari},
	journal={SciPost Phys.},
	volume={14},
	pages={007},
	year={2023},
	publisher={SciPost},
	doi={10.21468/SciPostPhys.14.1.007},
	url={https://scipost.org/10.21468/SciPostPhys.14.1.007},
}

@Article{Bhardwaj2025,
	title={{Generalized charges, part II: Non-invertible symmetries and the symmetry TFT}},
	author={Lakshya Bhardwaj and Sakura Schäfer-Nameki},
	journal={SciPost Phys.},
	volume={19},
	pages={098},
	year={2025},
	publisher={SciPost},
	doi={10.21468/SciPostPhys.19.4.098},
	url={https://scipost.org/10.21468/SciPostPhys.19.4.098},
}

@article{Bartsch2023-II,
    author = "Bartsch, Thomas and Bullimore, Mathew and Grigoletto, Andrea",
    title = "{Representation theory for categorical symmetries}",
    eprint = "2305.17165",
    archivePrefix = "arXiv",
    primaryClass = "hep-th",
    month = "5",
    year = "2023"
}

@Article{Tachikawa2020,
	title={{On gauging finite subgroups}},
	author={Yuji Tachikawa},
	journal={SciPost Phys.},
	volume={8},
	pages={015},
	year={2020},
	publisher={SciPost},
	doi={10.21468/SciPostPhys.8.1.015},
	url={https://scipost.org/10.21468/SciPostPhys.8.1.015},
}

@Article{Bhardwaj2023-II,
	title={{Unifying constructions of non-invertible symmetries}},
	author={Lakshya Bhardwaj and Sakura Schäfer-Nameki and Apoorv Tiwari},
	journal={SciPost Phys.},
	volume={15},
	pages={122},
	year={2023},
	publisher={SciPost},
	doi={10.21468/SciPostPhys.15.3.122},
	url={https://scipost.org/10.21468/SciPostPhys.15.3.122},
}

@Article{Bhardwaj2023-III,
	title={{Non-invertible symmetry webs}},
	author={Lakshya Bhardwaj and Lea E. Bottini and Sakura Schäfer-Nameki and Apoorv Tiwari},
	journal={SciPost Phys.},
	volume={15},
	pages={160},
	year={2023},
	publisher={SciPost},
	doi={10.21468/SciPostPhys.15.4.160},
	url={https://scipost.org/10.21468/SciPostPhys.15.4.160},
}

@article{SHS-non-invertible,
  title = {Noninvertible Global Symmetries in the Standard Model},
  author = {Choi, Yichul and Lam, Ho Tat and Shao, Shu-Heng},
  journal = {Phys. Rev. Lett.},
  volume = {129},
  issue = {16},
  pages = {161601},
  numpages = {6},
  year = {2022},
  month = {Oct},
  publisher = {American Physical Society},
  doi = {10.1103/PhysRevLett.129.161601},
  url = {https://link.aps.org/doi/10.1103/PhysRevLett.129.161601}
}

@Article{SHS-non-invertible-II,
author={Choi, Yichul
and Lam, Ho Tat
and Shao, Shu-Heng},
title={Non-invertible Gauss law and axions},
journal={Journal of High Energy Physics},
year={2023},
month={Sep},
day={12},
volume={2023},
number={9},
pages={67},
issn={1029-8479},
doi={10.1007/JHEP09(2023)067},
url={https://doi.org/10.1007/JHEP09(2023)067}
}

@article{Cordova-non-invertible,
  title = {Noninvertible Chiral Symmetry and Exponential Hierarchies},
  author = {C\'ordova, Clay and Ohmori, Kantaro},
  journal = {Phys. Rev. X},
  volume = {13},
  issue = {1},
  pages = {011034},
  numpages = {13},
  year = {2023},
  month = {Mar},
  publisher = {American Physical Society},
  doi = {10.1103/PhysRevX.13.011034},
  url = {https://link.aps.org/doi/10.1103/PhysRevX.13.011034}
}

@Article{Cordova2024-II,
author={C{\'o}rdova, Clay
and Hong, Sungwoo
and Wang, Lian-Tao},
title={Axion domain walls, small instantons, and non-invertible symmetry breaking},
journal={Journal of High Energy Physics},
year={2024},
month={May},
day={30},
volume={2024},
number={5},
pages={325},
issn={1029-8479},
doi={10.1007/JHEP05(2024)325},
url={https://doi.org/10.1007/JHEP05(2024)325}
}

@article{Callan-Harvey,
title = {Anomalies and fermion zero modes on strings and domain walls},
journal = {Nuclear Physics B},
volume = {250},
number = {1},
pages = {427-436},
year = {1985},
issn = {0550-3213},
doi = {https://doi.org/10.1016/0550-3213(85)90489-4},
url = {https://www.sciencedirect.com/science/article/pii/0550321385904894},
author = {C.G. Callan and J.A. Harvey}
}

@article{Naculich,
title = {Axionic strings: Covariant anomalies and bosonization of chiral zero modes},
journal = {Nuclear Physics B},
volume = {296},
number = {4},
pages = {837-867},
year = {1988},
issn = {0550-3213},
doi = {https://doi.org/10.1016/0550-3213(88)90400-2},
url = {https://www.sciencedirect.com/science/article/pii/0550321388904002},
author = {Stephen G. Naculich}
}

@article{Jackiw1975,
    author = "Jackiw, R.",
    title = "{Charge and Mass Spectrum of Quantum Solitons}",
    reportNumber = "MIT-CTP-510",
    journal = "Conf. Proc. C",
    volume = "750926",
    pages = "377--401",
    year = "1975"
}

@misc{harvey-anomalies,
      title={TASI 2003 Lectures on Anomalies}, 
      author={Jeffrey A. Harvey},
      year={2005},
      eprint={hep-th/0509097},
      archivePrefix={arXiv},
      primaryClass={hep-th},
      url={https://arxiv.org/abs/hep-th/0509097}, 
}

@Article{anomalies-coupling-constants,
	title={{Anomalies in the space of coupling constants and their dynamical applications II}},
	author={Clay Córdova and Daniel S. Freed and Ho Tat Lam and Nathan Seiberg},
	journal={SciPost Phys.},
	volume={8},
	pages={002},
	year={2020},
	publisher={SciPost},
	doi={10.21468/SciPostPhys.8.1.002},
	url={https://scipost.org/10.21468/SciPostPhys.8.1.002},
}

@Article{Gaiotto2017,
author={Gaiotto, Davide
and Kapustin, Anton
and Komargodski, Zohar
and Seiberg, Nathan},
title={Theta, time reversal and temperature},
journal={Journal of High Energy Physics},
year={2017},
month={May},
day={17},
volume={2017},
number={5},
pages={91},
issn={1029-8479},
doi={10.1007/JHEP05(2017)091},
url={https://doi.org/10.1007/JHEP05(2017)091}
}

@Article{Seiberg-line-operators,
author={Aharony, Ofer
and Seiberg, Nathan
and Tachikawa, Yuji},
title={Reading between the lines of four-dimensional gauge theories},
journal={Journal of High Energy Physics},
year={2013},
month={Aug},
day={26},
volume={2013},
number={8},
pages={115},
issn={1029-8479},
doi={10.1007/JHEP08(2013)115},
url={https://doi.org/10.1007/JHEP08(2013)115}
}

@article{cheung2024,
    author = "Cheung, Clifford and Derda, Maria and Kim, Joon-Hwi and Nevoa, Vinicius and Rothstein, Ira and Shah, Nabha",
    title = "{Generalized symmetry in dynamical gravity}",
    eprint = "2403.01837",
    archivePrefix = "arXiv",
    primaryClass = "hep-th",
    reportNumber = "CALT-TH 2024-009",
    doi = "10.1007/JHEP10(2024)007",
    journal = "JHEP",
    volume = "10",
    pages = "007",
    year = "2024"
}

@Article{Aguilera,
	title={{Non-invertible defects in 5d, boundaries and holography}},
	author={Jeremias Aguilera Damia and Riccardo Argurio and Eduardo Garcia-Valdecasas},
	journal={SciPost Phys.},
	volume={14},
	pages={067},
	year={2023},
	publisher={SciPost},
	doi={10.21468/SciPostPhys.14.4.067},
	url={https://scipost.org/10.21468/SciPostPhys.14.4.067},
}

@article{Yokokura,
    author = "Yokokura, Ryo",
    title = "{Non-invertible symmetries in axion electrodynamics}",
    eprint = "2212.05001",
    archivePrefix = "arXiv",
    primaryClass = "hep-th",
    reportNumber = "KEK-TH-2481",
    month = "12",
    year = "2022"
}

@inproceedings{Sundrum,
    author = "Sundrum, Raman",
    title = "{Tasi 2004 lectures: To the fifth dimension and back}",
    booktitle = "{Theoretical Advanced Study Institute in Elementary Particle Physics}: {Physics in D $\geqq$ 4}",
    eprint = "hep-th/0508134",
    archivePrefix = "arXiv",
    pages = "585--630",
    month = "8",
    year = "2005"
}

@article{Reece,
    author = "Reece, Matthew",
    title = "{TASI Lectures: (No) Global Symmetries to Axion Physics}",
    eprint = "2304.08512",
    archivePrefix = "arXiv",
    primaryClass = "hep-ph",
    doi = "10.22323/1.439.0008",
    journal = "PoS",
    volume = "TASI2022",
    pages = "008",
    year = "2024"
}

@article{5daxionYangMillsI,
  title = {QCD Axion from a Higher Dimensional Gauge Field Theory},
  author = {Choi, Kiwoon},
  journal = {Phys. Rev. Lett.},
  volume = {92},
  issue = {10},
  pages = {101602},
  numpages = {4},
  year = {2004},
  month = {Mar},
  publisher = {American Physical Society},
  doi = {10.1103/PhysRevLett.92.101602},
  url = {https://link.aps.org/doi/10.1103/PhysRevLett.92.101602}
}

@article{5daxionYangMillsII,
  title = {Note on the strong $CP$ problem from a 5-dimensional perspective},
  author = {Grzadkowski, Bohdan and Wudka, Jos\'e},
  journal = {Phys. Rev. D},
  volume = {77},
  issue = {9},
  pages = {096004},
  numpages = {6},
  year = {2008},
  month = {May},
  publisher = {American Physical Society},
  doi = {10.1103/PhysRevD.77.096004},
  url = {https://link.aps.org/doi/10.1103/PhysRevD.77.096004}
}

@Article{Bonetti2013,
author={Bonetti, Federico
and Grimm, Thomas W.
and Hohenegger, Stefan},
title={One-loop Chern-Simons terms in five dimensions},
journal={Journal of High Energy Physics},
year={2013},
month={Jul},
day={08},
volume={2013},
number={7},
pages={43},
issn={1029-8479},
doi={10.1007/JHEP07(2013)043},
url={https://doi.org/10.1007/JHEP07(2013)043}
}

@article{5dGG,
title = {A Toy Model of the M5-Brane: Anomalies of Monopole Strings in Five Dimensions},
journal = {Annals of Physics},
volume = {301},
number = {1},
pages = {1-21},
year = {2002},
issn = {0003-4916},
doi = {https://doi.org/10.1006/aphy.2002.6294},
url = {https://www.sciencedirect.com/science/article/pii/S0003491602962949},
author = {Alexey Boyarsky and Jeffrey A. Harvey and Oleg Ruchayskiy}
}

@Article{Barkeshli,
	title={{Higher-group symmetry in finite gauge theory and stabilizer codes}},
	author={Maissam Barkeshli and Yu-An Chen and Po-Shen Hsin and Ryohei Kobayashi},
	journal={SciPost Phys.},
	volume={16},
	pages={089},
	year={2024},
	publisher={SciPost},
	doi={10.21468/SciPostPhys.16.4.089},
	url={https://scipost.org/10.21468/SciPostPhys.16.4.089},
}

@misc{tong2016lecturesquantumhalleffect,
      title={Lectures on the Quantum Hall Effect}, 
      author={David Tong},
      year={2016},
      eprint={1606.06687},
      archivePrefix={arXiv},
      primaryClass={hep-th},
      url={https://arxiv.org/abs/1606.06687}, 
}

@misc{tonggt,
  author        = {David Tong},
  title         = {Gauge theory},
  year          = {2018},
  publisher={DAMTP Cambridge},
  url={https://www.damtp.cam.ac.uk/user/tong/gaugetheory.html}
}

@Article{Tong2017,
author={Tong, David},
title={Line operators in the Standard Model},
journal={Journal of High Energy Physics},
year={2017},
month={Jul},
day={20},
volume={2017},
number={7},
pages={104},
issn={1029-8479},
doi={10.1007/JHEP07(2017)104},
url={https://doi.org/10.1007/JHEP07(2017)104}
}

@Article{Hidaka2021,
author={Hidaka, Yoshimasa
and Nitta, Muneto
and Yokokura, Ryo},
title={Global 3-group symmetry and 't Hooft anomalies in axion electrodynamics},
journal={Journal of High Energy Physics},
year={2021},
month={Jan},
day={27},
volume={2021},
number={1},
pages={173},
issn={1029-8479},
doi={10.1007/JHEP01(2021)173},
url={https://doi.org/10.1007/JHEP01(2021)173}
}

@article{Hidaka2020,
title = {Higher-form symmetries and 3-group in axion electrodynamics},
journal = {Physics Letters B},
volume = {808},
pages = {135672},
year = {2020},
issn = {0370-2693},
doi = {https://doi.org/10.1016/j.physletb.2020.135672},
url = {https://www.sciencedirect.com/science/article/pii/S0370269320304755},
author = {Yoshimasa Hidaka and Muneto Nitta and Ryo Yokokura}
}

@Article{Gaiotto2018,
author={Gaiotto, Davide
and Komargodski, Zohar
and Seiberg, Nathan},
title={Time-reversal breaking in QCD4, walls, and dualities in 2 + 1 dimensions},
journal={Journal of High Energy Physics},
year={2018},
month={Jan},
day={23},
volume={2018},
number={1},
pages={110},
issn={1029-8479},
doi={10.1007/JHEP01(2018)110},
url={https://doi.org/10.1007/JHEP01(2018)110}
}

@Article{Seiberg2019,
	title={{Comments on one-form global symmetries and their gauging in 3d and 4d}},
	author={Po-Shen Hsin and Ho Tat Lam and Nathan Seiberg},
	journal={SciPost Phys.},
	volume={6},
	pages={039},
	year={2019},
	publisher={SciPost},
	doi={10.21468/SciPostPhys.6.3.039},
	url={https://scipost.org/10.21468/SciPostPhys.6.3.039},
}

@Article{ReeceTwist,
author={Reece, Matthew},
title={Axion-gauge coupling quantization with a twist},
journal={Journal of High Energy Physics},
year={2023},
month={Oct},
day={19},
volume={2023},
number={10},
pages={116},
issn={1029-8479},
doi={10.1007/JHEP10(2023)116},
url={https://doi.org/10.1007/JHEP10(2023)116}
}

@article{Olive1978,
    author = "Olive, David I.",
    title = "{MAGNETIC MONOPOLES}",
    reportNumber = "ICTP-77-78-20",
    doi = "10.1016/0370-1573(79)90105-4",
    journal = "Phys. Rept.",
    volume = "49",
    pages = "165--172",
    year = "1979"
}

@article{Chern-Weil-II,
    author = "Garc{\'\i}a-Valdecasas, Eduardo and Reece, Matthew and Suzuki, Motoo",
    title = "{Monopole breaking of Chern-Weil symmetries}",
    eprint = "2408.00067",
    archivePrefix = "arXiv",
    primaryClass = "hep-th",
    doi = "10.21468/SciPostPhys.18.5.162",
    journal = "SciPost Phys.",
    volume = "18",
    number = "5",
    pages = "162",
    year = "2025"
}

@article{MonopoleLoops,
  title = {Axion Mass from Magnetic Monopole Loops},
  author = {Fan, JiJi and Fraser, Katherine and Reece, Matthew and Stout, John},
  journal = {Phys. Rev. Lett.},
  volume = {127},
  issue = {13},
  pages = {131602},
  numpages = {6},
  year = {2021},
  month = {Sep},
  publisher = {American Physical Society},
  doi = {10.1103/PhysRevLett.127.131602},
  url = {https://link.aps.org/doi/10.1103/PhysRevLett.127.131602}
}

@Article{Fukuda2021,
author={Fukuda, Hajime
and Yonekura, Kazuya},
title={Witten effect, anomaly inflow, and charge teleportation},
journal={Journal of High Energy Physics},
year={2021},
month={Jan},
day={20},
volume={2021},
number={1},
pages={119},
issn={1029-8479},
doi={10.1007/JHEP01(2021)119},
url={https://doi.org/10.1007/JHEP01(2021)119}
}

@article{craig2024highqualityaxionshigherformsymmetries,
    author = "Craig, Nathaniel and Kongsore, Marius",
    title = "{High-quality axions from higher-form symmetries in extra dimensions}",
    eprint = "2408.10295",
    archivePrefix = "arXiv",
    primaryClass = "hep-ph",
    doi = "10.1103/PhysRevD.111.015047",
    journal = "Phys. Rev. D",
    volume = "111",
    number = "1",
    pages = "015047",
    year = "2025"
}

@book{Wen_QFT,
    author = "Wen, Xiao-Gang",
    title = "{Quantum Field Theory of Many-Body Systems}",
    doi = "10.1093/acprof:oso/9780199227259.001.0001",
    isbn = "978-0-19-171301-9, 978-0-19-922725-9",
    publisher = "Oxford University Press",
    month = "9",
    year = "2007"
}

@Article{QCD-SPT,
	title={{Higgs-confinement transitions in QCD from symmetry protected topological phases}},
	author={Thomas T. Dumitrescu and Po-Shen Hsin},
	journal={SciPost Phys.},
	volume={17},
	pages={093},
	year={2024},
	publisher={SciPost},
	doi={10.21468/SciPostPhys.17.3.093},
	url={https://scipost.org/10.21468/SciPostPhys.17.3.093},
}

@Article{Heidenreich2021,
author={Heidenreich, Ben
and Reece, Matthew
and Rudelius, Tom},
title={The Weak Gravity Conjecture and axion strings},
journal={Journal of High Energy Physics},
year={2021},
month={Nov},
day={02},
volume={2021},
number={11},
pages={4},
issn={1029-8479},
doi={10.1007/JHEP11(2021)004},
url={https://doi.org/10.1007/JHEP11(2021)004}
}

@Article{Gaiotto2015,
author={Gaiotto, Davide
and Kapustin, Anton
and Seiberg, Nathan
and Willett, Brian},
title={Generalized global symmetries},
journal={Journal of High Energy Physics},
year={2015},
month={Feb},
day={26},
volume={2015},
number={2},
pages={172},
issn={1029-8479},
doi={10.1007/JHEP02(2015)172},
url={https://doi.org/10.1007/JHEP02(2015)172}
}

@article{Verlinde1988,
    author = "Verlinde, Erik P.",
    title = "{Fusion Rules and Modular Transformations in 2D Conformal Field Theory}",
    reportNumber = "THU-88/17",
    doi = "10.1016/0550-3213(88)90603-7",
    journal = "Nucl. Phys. B",
    volume = "300",
    pages = "360--376",
    year = "1988"
}

@article{Petkova2001,
title = {Generalised twisted partition functions},
journal = {Physics Letters B},
volume = {504},
number = {1},
pages = {157-164},
year = {2001},
issn = {0370-2693},
doi = {https://doi.org/10.1016/S0370-2693(01)00276-3},
url = {https://www.sciencedirect.com/science/article/pii/S0370269301002763},
author = {V.B. Petkova and J.-B. Zuber}
}

@article{Fuchs2002,
title = {TFT construction of RCFT correlators I: partition functions},
journal = {Nuclear Physics B},
volume = {646},
number = {3},
pages = {353-497},
year = {2002},
issn = {0550-3213},
doi = {https://doi.org/10.1016/S0550-3213(02)00744-7},
url = {https://www.sciencedirect.com/science/article/pii/S0550321302007447},
author = {Jürgen Fuchs and Ingo Runkel and Christoph Schweigert}
}

@article{Frohlich2004,
  title = {Kramers-Wannier Duality from Conformal Defects},
  author = {Fr\"ohlich, J\"urg and Fuchs, J\"urgen and Runkel, Ingo and Schweigert, Christoph},
  journal = {Phys. Rev. Lett.},
  volume = {93},
  issue = {7},
  pages = {070601},
  numpages = {4},
  year = {2004},
  month = {Aug},
  publisher = {American Physical Society},
  doi = {10.1103/PhysRevLett.93.070601},
  url = {https://link.aps.org/doi/10.1103/PhysRevLett.93.070601}
}

@article{Frohlich2007,
title = {Duality and defects in rational conformal field theory},
journal = {Nuclear Physics B},
volume = {763},
number = {3},
pages = {354-430},
year = {2007},
issn = {0550-3213},
doi = {https://doi.org/10.1016/j.nuclphysb.2006.11.017},
url = {https://www.sciencedirect.com/science/article/pii/S0550321306009187},
author = {Jürg Fröhlich and Jürgen Fuchs and Ingo Runkel and Christoph Schweigert}
}

@article{Aasen2016,
doi = {10.1088/1751-8113/49/35/354001},
url = {https://dx.doi.org/10.1088/1751-8113/49/35/354001},
year = {2016},
month = {aug},
publisher = {IOP Publishing},
volume = {49},
number = {35},
pages = {354001},
author = {David Aasen and Roger S K Mong and Paul Fendley},
title = {Topological defects on the lattice: I. The Ising model},
journal = {Journal of Physics A: Mathematical and Theoretical}
}

@Article{Bhardwaj2018,
author={Bhardwaj, Lakshya
and Tachikawa, Yuji},
title={On finite symmetries and their gauging in two dimensions},
journal={Journal of High Energy Physics},
year={2018},
month={Mar},
day={29},
volume={2018},
number={3},
pages={189},
issn={1029-8479},
doi={10.1007/JHEP03(2018)189},
url={https://doi.org/10.1007/JHEP03(2018)189}
}

@Article{Chang2019,
author={Chang, Chi-Ming
and Lin, Ying-Hsuan
and Shao, Shu-Heng
and Wang, Yifan
and Yin, Xi},
title={Topological defect lines and renormalization group flows in two dimensions},
journal={Journal of High Energy Physics},
year={2019},
month={Jan},
day={03},
volume={2019},
number={1},
pages={26},
issn={1029-8479},
doi={10.1007/JHEP01(2019)026},
url={https://doi.org/10.1007/JHEP01(2019)026}
}

@article{Lin2021,
doi = {10.1088/1751-8121/abd69e},
url = {https://dx.doi.org/10.1088/1751-8121/abd69e},
year = {2021},
month = {jan},
publisher = {IOP Publishing},
volume = {54},
number = {6},
pages = {065201},
author = {Ying-Hsuan Lin and Shu-Heng Shao},
title = {Duality defect of the monster CFT},
journal = {Journal of Physics A: Mathematical and Theoretical}
}

@Article{Jensen2018,
author={Jensen, Kristan
and Shaverin, Evgeny
and Yarom, Amos},
title={'t Hooft anomalies and boundaries},
journal={Journal of High Energy Physics},
year={2018},
month={Jan},
day={18},
volume={2018},
number={1},
pages={85},
issn={1029-8479},
doi={10.1007/JHEP01(2018)085},
url={https://doi.org/10.1007/JHEP01(2018)085}
}

@Article{Thorngren2021,
author={Thorngren, Ryan
and Wang, Yifan},
title={Anomalous symmetries end at the boundary},
journal={Journal of High Energy Physics},
year={2021},
month={Sep},
day={03},
volume={2021},
number={9},
pages={17},
issn={1029-8479},
doi={10.1007/JHEP09(2021)017},
url={https://doi.org/10.1007/JHEP09(2021)017}
}

@Article{Thorngren2024,
author={Thorngren, Ryan
and Wang, Yifan},
title={Fusion category symmetry. Part I. Anomaly in-flow and gapped phases},
journal={Journal of High Energy Physics},
year={2024},
month={Apr},
day={24},
volume={2024},
number={4},
pages={132},
issn={1029-8479},
doi={10.1007/JHEP04(2024)132},
url={https://doi.org/10.1007/JHEP04(2024)132}
}

@article{aasen2020topologicaldefectslatticedualities,
    author = "Aasen, David and Fendley, Paul and Mong, Roger S. K.",
    title = "{Topological Defects on the Lattice: Dualities and Degeneracies}",
    eprint = "2008.08598",
    archivePrefix = "arXiv",
    primaryClass = "cond-mat.stat-mech",
    month = "8",
    year = "2020"
}

@article{Sharpe2023,
   title={Topological operators, noninvertible symmetries and decomposition},
   volume={27},
   ISSN={1095-0753},
   url={http://dx.doi.org/10.4310/ATMP.2023.v27.n8.a2},
   DOI={10.4310/atmp.2023.v27.n8.a2},
   number={8},
   journal={Advances in Theoretical and Mathematical Physics},
   publisher={International Press of Boston},
   author={Sharpe, Eric},
   year={2023},
   pages={2319–2407} }

@Article{Huang2021,
author={Huang, Tzu-Chen
and Lin, Ying-Hsuan
and Seifnashri, Sahand},
title={Construction of two-dimensional topological field theories with non-invertible symmetries},
journal={Journal of High Energy Physics},
year={2021},
month={Dec},
day={07},
volume={2021},
number={12},
pages={28},
issn={1029-8479},
doi={10.1007/JHEP12(2021)028},
url={https://doi.org/10.1007/JHEP12(2021)028}
}

@Article{Chang2023,
	title={{Topological defect lines in two dimensional fermionic CFTs}},
	author={Chi-Ming Chang and Jin Chen and Fengjun Xu},
	journal={SciPost Phys.},
	volume={15},
	pages={216},
	year={2023},
	publisher={SciPost},
	doi={10.21468/SciPostPhys.15.5.216},
	url={https://scipost.org/10.21468/SciPostPhys.15.5.216},
}

@Article{Lin2023,
author={Lin, Ying-Hsuan
and Okada, Masaki
and Seifnashri, Sahand
and Tachikawa, Yuji},
title={Asymptotic density of states in 2d CFTs with non-invertible symmetries},
journal={Journal of High Energy Physics},
year={2023},
month={Mar},
day={14},
volume={2023},
number={3},
pages={94},
issn={1029-8479},
doi={10.1007/JHEP03(2023)094},
url={https://doi.org/10.1007/JHEP03(2023)094}
}

@inproceedings{shao2024whatsundonetasilectures,
    author = "Shao, Shu-Heng",
    title = "{What's Done Cannot Be Undone: TASI Lectures on Non-Invertible Symmetries}",
    booktitle = "{Theoretical Advanced Study Institute in Elementary Particle Physics 2023}: {Aspects of Symmetry}",
    eprint = "2308.00747",
    archivePrefix = "arXiv",
    primaryClass = "hep-th",
    reportNumber = "YITP-SB-2023-19",
    month = "8",
    year = "2023"
}

@article{Kapustin2017,
    author = "Kapustin, Anton and Thorngren, Ryan",
    title = "{Higher Symmetry and Gapped Phases of Gauge Theories}",
    eprint = "1309.4721",
    archivePrefix = "arXiv",
    primaryClass = "hep-th",
    doi = "10.1007/978-3-319-59939-7_5",
    journal = "Prog. Math.",
    volume = "324",
    pages = "177--202",
    year = "2017"
}

@Article{Cordova2019,
author={C{\'o}rdova, Clay
and Dumitrescu, Thomas T.
and Intriligator, Kenneth},
title={Exploring 2-group global symmetries},
journal={Journal of High Energy Physics},
year={2019},
month={Feb},
day={27},
volume={2019},
number={2},
pages={184},
issn={1029-8479},
doi={10.1007/JHEP02(2019)184},
url={https://doi.org/10.1007/JHEP02(2019)184}
}

@article{Koren2022,
    author = "Cordova, Clay and Koren, Seth",
    title = "{Higher Flavor Symmetries in the Standard Model}",
    eprint = "2212.13193",
    archivePrefix = "arXiv",
    primaryClass = "hep-ph",
    doi = "10.1002/andp.202300031",
    journal = "Annalen Phys.",
    volume = "535",
    number = "8",
    pages = "2300031",
    year = "2023"
}

@Article{Benini2019,
author={Benini, Francesco
and C{\'o}rdova, Clay
and Hsin, Po-Shen},
title={On 2-group global symmetries and their anomalies},
journal={Journal of High Energy Physics},
year={2019},
month={Mar},
day={21},
volume={2019},
number={3},
pages={118},
issn={1029-8479},
doi={10.1007/JHEP03(2019)118},
url={https://doi.org/10.1007/JHEP03(2019)118}
}

@article{Barkeshli2018,
  title = {Time-reversal and spatial-reflection symmetry localization anomalies in (2+1)-dimensional topological phases of matter},
  author = {Barkeshli, Maissam and Cheng, Meng},
  journal = {Phys. Rev. B},
  volume = {98},
  issue = {11},
  pages = {115129},
  numpages = {19},
  year = {2018},
  month = {Sep},
  publisher = {American Physical Society},
  doi = {10.1103/PhysRevB.98.115129},
  url = {https://link.aps.org/doi/10.1103/PhysRevB.98.115129}
}

@article{Fidkowski2017,
  title = {Realizing anomalous anyonic symmetries at the surfaces of three-dimensional gauge theories},
  author = {Fidkowski, Lukasz and Vishwanath, Ashvin},
  journal = {Phys. Rev. B},
  volume = {96},
  issue = {4},
  pages = {045131},
  numpages = {14},
  year = {2017},
  month = {Jul},
  publisher = {American Physical Society},
  doi = {10.1103/PhysRevB.96.045131},
  url = {https://link.aps.org/doi/10.1103/PhysRevB.96.045131}
}

@Article{Hsin2020,
author={Hsin, Po-Shen
and Turzillo, Alex},
title={Symmetry-enriched quantum spin liquids in (3 + 1)d},
journal={Journal of High Energy Physics},
year={2020},
month={Sep},
day={02},
volume={2020},
number={9},
pages={22},
issn={1029-8479},
doi={10.1007/JHEP09(2020)022},
url={https://doi.org/10.1007/JHEP09(2020)022}
}

@Article{Tanizaki2020,
author={Tanizaki, Yuya
and {\"U}nsal, Mithat},
title={Modified instanton sum in QCD and higher-groups},
journal={Journal of High Energy Physics},
year={2020},
month={Mar},
day={23},
volume={2020},
number={3},
pages={123},
issn={1029-8479},
doi={10.1007/JHEP03(2020)123},
url={https://doi.org/10.1007/JHEP03(2020)123}
}

@Article{Iqbal2023,
	title={{2-group global symmetries, hydrodynamics and holography}},
	author={Nabil Iqbal and Napat Poovuttikul},
	journal={SciPost Phys.},
	volume={15},
	pages={063},
	year={2023},
	publisher={SciPost},
	doi={10.21468/SciPostPhys.15.2.063},
	url={https://scipost.org/10.21468/SciPostPhys.15.2.063},
}

@Article{Cordova2021,
author={C{\'o}rdova, Clay
and Dumitrescu, Thomas T.
and Intriligator, Kenneth},
title={2-Group global symmetries and anomalies in six-dimensional quantum field theories},
journal={Journal of High Energy Physics},
year={2021},
month={Apr},
day={26},
volume={2021},
number={4},
pages={252},
issn={1029-8479},
doi={10.1007/JHEP04(2021)252},
url={https://doi.org/10.1007/JHEP04(2021)252}
}

@Article{Apruzzi2022,
	title={{2-Group symmetries and their classification in 6d}},
	author={Fabio Apruzzi and Lakshya Bhardwaj and Dewi S. W. Gould and Sakura Schäfer-Nameki},
	journal={SciPost Phys.},
	volume={12},
	pages={098},
	year={2022},
	publisher={SciPost},
	doi={10.21468/SciPostPhys.12.3.098},
	url={https://scipost.org/10.21468/SciPostPhys.12.3.098},
}

@Article{Hsin2022,
	title={{Exotic invertible phases with higher-group symmetries}},
	author={Po-Shen Hsin and Wenjie Ji and Chao-Ming Jian},
	journal={SciPost Phys.},
	volume={12},
	pages={052},
	year={2022},
	publisher={SciPost},
	doi={10.21468/SciPostPhys.12.2.052},
	url={https://scipost.org/10.21468/SciPostPhys.12.2.052},
}

@article{DeWolfe2021,
  title = {Generalized symmetries and 2-groups via electromagnetic duality in $\mathrm{AdS}/\mathrm{CFT}$},
  author = {DeWolfe, Oliver and Higginbotham, Kenneth},
  journal = {Phys. Rev. D},
  volume = {103},
  issue = {2},
  pages = {026011},
  numpages = {13},
  year = {2021},
  month = {Jan},
  publisher = {American Physical Society},
  doi = {10.1103/PhysRevD.103.026011},
  url = {https://link.aps.org/doi/10.1103/PhysRevD.103.026011}
}

@Article{Delmastro2023,
	title={{Anomalies and symmetry fractionalization}},
	author={Diego Delmastro and Jaume Gomis and Po-Shen Hsin and Zohar Komargodski},
	journal={SciPost Phys.},
	volume={15},
	pages={079},
	year={2023},
	publisher={SciPost},
	doi={10.21468/SciPostPhys.15.3.079},
	url={https://scipost.org/10.21468/SciPostPhys.15.3.079},
}

@Article{BarkeshliII,
	title={{Higher-group symmetry of (3+1)D fermionic $\mathbb{Z}_2$ gauge theory: Logical CCZ, CS, and T gates from higher symmetry}},
	author={Maissam Barkeshli and Po-Shen Hsin and Ryohei Kobayashi},
	journal={SciPost Phys.},
	volume={16},
	pages={122},
	year={2024},
	publisher={SciPost},
	doi={10.21468/SciPostPhys.16.5.122},
	url={https://scipost.org/10.21468/SciPostPhys.16.5.122},
}

@article{seifnashri2024clusterstatenoninvertiblesymmetry,
    author = "Seifnashri, Sahand and Shao, Shu-Heng",
    title = "{Cluster State as a Noninvertible Symmetry-Protected Topological Phase}",
    eprint = "2404.01369",
    archivePrefix = "arXiv",
    primaryClass = "cond-mat.str-el",
    reportNumber = "YITP-SB-2024-05",
    doi = "10.1103/PhysRevLett.133.116601",
    journal = "Phys. Rev. Lett.",
    volume = "133",
    number = "11",
    pages = "116601",
    year = "2024"
}

@article{Cordova2024,
    author = "Cordova, Clay and Costa, Davi Bastos and Hsin, Po-Shen",
    title = "{Non-invertible symmetries in finite-group gauge theory}",
    eprint = "2407.07964",
    archivePrefix = "arXiv",
    primaryClass = "cond-mat.str-el",
    doi = "10.21468/SciPostPhys.18.1.019",
    journal = "SciPost Phys.",
    volume = "18",
    number = "1",
    pages = "019",
    year = "2025"
}

@Article{Roumpedakis2023,
author={Roumpedakis, Konstantinos
and Seifnashri, Sahand
and Shao, Shu-Heng},
title={Higher Gauging and Non-invertible Condensation Defects},
journal={Communications in Mathematical Physics},
year={2023},
month={Aug},
day={01},
volume={401},
number={3},
pages={3043-3107},
issn={1432-0916},
doi={10.1007/s00220-023-04706-9},
url={https://doi.org/10.1007/s00220-023-04706-9}
}

@Article{Damia2023,
author={Damia, Jeremias Aguilera
and Argurio, Riccardo
and Tizzano, Luigi},
title={Continuous generalized symmetries in three dimensions},
journal={Journal of High Energy Physics},
year={2023},
month={May},
day={19},
volume={2023},
number={5},
pages={164},
issn={1029-8479},
doi={10.1007/JHEP05(2023)164},
url={https://doi.org/10.1007/JHEP05(2023)164}
}

@article{choi2024noninvertiblehigherformsymmetries21d,
    author = "Choi, Yichul and Sanghavi, Yaman and Shao, Shu-Heng and Zheng, Yunqin",
    title = "{Non-invertible and higher-form symmetries in 2+1d lattice gauge theories}",
    eprint = "2405.13105",
    archivePrefix = "arXiv",
    primaryClass = "cond-mat.str-el",
    doi = "10.21468/SciPostPhys.18.1.008",
    journal = "SciPost Phys.",
    volume = "18",
    number = "1",
    pages = "008",
    year = "2025"
}

@Article{Argurio2023,
author={Argurio, Riccardo
and Vandepopeliere, Romain},
title={When $\mathbb{Z}_2$ one-form symmetry leads to non-invertible axial symmetries},
journal={Journal of High Energy Physics},
year={2023},
month={Aug},
day={30},
volume={2023},
number={8},
pages={205},
issn={1029-8479},
doi={10.1007/JHEP08(2023)205},
url={https://doi.org/10.1007/JHEP08(2023)205}
}

@Article{Bhardwaj2024,
	title={{Generalized symmetries and anomalies of 3d $\mathcal{N}=4$ SCFTs}},
	author={Lakshya Bhardwaj and Mathew Bullimore and Andrea E. V. Ferrari and Sakura Schäfer-Nameki},
	journal={SciPost Phys.},
	volume={16},
	pages={080},
	year={2024},
	publisher={SciPost},
	doi={10.21468/SciPostPhys.16.3.080},
	url={https://scipost.org/10.21468/SciPostPhys.16.3.080},
}

@article{Kang2024,
    author = "Kang, Monica Jinwoo and Lawrie, Craig and Lee, Ki-Hong and Song, Jaewon",
    title = "{Generalized symmetry constraints on deformed 4D SCFTs}",
    eprint = "2408.14532",
    archivePrefix = "arXiv",
    primaryClass = "hep-th",
    reportNumber = "DESY-24-125",
    doi = "10.1103/PhysRevD.111.086028",
    journal = "Phys. Rev. D",
    volume = "111",
    number = "8",
    pages = "086028",
    year = "2025"
}

@article{Heckman2023,
author = {Heckman, Jonathan J. and Hübner, Max and Torres, Ethan and Zhang, Hao Y.},
title = {The Branes Behind Generalized Symmetry Operators},
journal = {Fortschritte der Physik},
volume = {71},
number = {1},
pages = {2200180},
doi = {https://doi.org/10.1002/prop.202200180},
url = {https://onlinelibrary.wiley.com/doi/abs/10.1002/prop.202200180},
eprint = {https://onlinelibrary.wiley.com/doi/pdf/10.1002/prop.202200180},
year = {2023}
}

@article{I1,
    author = "Koide, Masataka and Nagoya, Yuta and Yamaguchi, Satoshi",
    title = "{Non-invertible topological defects in 4-dimensional $\mathbb {Z}_2$ pure lattice gauge theory}",
    eprint = "2109.05992",
    archivePrefix = "arXiv",
    primaryClass = "hep-th",
    reportNumber = "OU-HET 1105",
    doi = "10.1093/ptep/ptab145",
    journal = "PTEP",
    volume = "2022",
    number = "1",
    pages = "013B03",
    year = "2022"
}

@article{I2,
  title = {Noninvertible duality defects in $3+1$ dimensions},
  author = {Choi, Yichul and C\'ordova, Clay and Hsin, Po-Shen and Lam, Ho Tat and Shao, Shu-Heng},
  journal = {Phys. Rev. D},
  volume = {105},
  issue = {12},
  pages = {125016},
  numpages = {19},
  year = {2022},
  month = {Jun},
  publisher = {American Physical Society},
  doi = {10.1103/PhysRevD.105.125016},
  url = {https://link.aps.org/doi/10.1103/PhysRevD.105.125016}
}

@article{I3,
  title = {Kramers-Wannier-like Duality Defects in $(3+1)D$ Gauge Theories},
  author = {Kaidi, Justin and Ohmori, Kantaro and Zheng, Yunqin},
  journal = {Phys. Rev. Lett.},
  volume = {128},
  issue = {11},
  pages = {111601},
  numpages = {6},
  year = {2022},
  month = {Mar},
  publisher = {American Physical Society},
  doi = {10.1103/PhysRevLett.128.111601},
  url = {https://link.aps.org/doi/10.1103/PhysRevLett.128.111601}
}

@Article{I4,
author={C{\'o}rdova, Clay
and Ohmori, Kantaro
and Rudelius, Tom},
title={Generalized symmetry breaking scales and weak gravity conjectures},
journal={Journal of High Energy Physics},
year={2022},
month={Nov},
day={28},
volume={2022},
number={11},
pages={154},
issn={1029-8479},
doi={10.1007/JHEP11(2022)154},
url={https://doi.org/10.1007/JHEP11(2022)154}
}

@Article{I5,
author={Arias-Tamargo, Guillermo
and Rodr{\'i}guez-G{\'o}mez, Diego},
title={Non-invertible symmetries from discrete gauging and completeness of the spectrum},
journal={Journal of High Energy Physics},
year={2023},
month={Apr},
day={20},
volume={2023},
number={4},
pages={93},
issn={1029-8479},
doi={10.1007/JHEP04(2023)093},
url={https://doi.org/10.1007/JHEP04(2023)093}
}

@Article{I6,
author={Kaidi, Justin
and Zafrir, Gabi
and Zheng, Yunqin},
title={Non-invertible symmetries of $\mathcal{N}=4$ SYM and twisted compactification},
journal={Journal of High Energy Physics},
year={2022},
month={Aug},
day={03},
volume={2022},
number={8},
pages={53},
issn={1029-8479},
doi={10.1007/JHEP08(2022)053},
url={https://doi.org/10.1007/JHEP08(2022)053}
}

@article{I7,
    author = "Bashmakov, Vladimir and Del Zotto, Michele and Hasan, Azeem",
    title = "{On the 6d origin of non-invertible symmetries in 4d}",
    eprint = "2206.07073",
    archivePrefix = "arXiv",
    primaryClass = "hep-th",
    doi = "10.1007/JHEP09(2023)161",
    journal = "JHEP",
    volume = "09",
    pages = "161",
    year = "2023"
}

@article{I8,
  title = {Noninvertible Time-Reversal Symmetry},
  author = {Choi, Yichul and Lam, Ho Tat and Shao, Shu-Heng},
  journal = {Phys. Rev. Lett.},
  volume = {130},
  issue = {13},
  pages = {131602},
  numpages = {9},
  year = {2023},
  month = {Mar},
  publisher = {American Physical Society},
  doi = {10.1103/PhysRevLett.130.131602},
  url = {https://link.aps.org/doi/10.1103/PhysRevLett.130.131602}
}

@article{I9,
author = {Bhardwaj, Lakshya and Schäfer-Nameki, Sakura and Wu, Jingxiang},
title = {Universal Non-Invertible Symmetries},
journal = {Fortschritte der Physik},
volume = {70},
number = {11},
pages = {2200143},
doi = {https://doi.org/10.1002/prop.202200143},
url = {https://onlinelibrary.wiley.com/doi/abs/10.1002/prop.202200143},
eprint = {https://onlinelibrary.wiley.com/doi/pdf/10.1002/prop.202200143},
year = {2022}
}

@Article{I10,
	title={{Non-invertible symmetries and higher representation theory I}},
	author={Thomas Bartsch and Mathew Bullimore and Andrea E. V. Ferrari and Jamie Pearson},
	journal={SciPost Phys.},
	volume={17},
	pages={015},
	year={2024},
	publisher={SciPost},
	doi={10.21468/SciPostPhys.17.1.015},
	url={https://scipost.org/10.21468/SciPostPhys.17.1.015},
}

@article{I11,
  title = {Noninvertible Symmetries from Holography and Branes},
  author = {Apruzzi, Fabio and Bah, Ibrahima and Bonetti, Federico and Sch\"afer-Nameki, Sakura},
  journal = {Phys. Rev. Lett.},
  volume = {130},
  issue = {12},
  pages = {121601},
  numpages = {6},
  year = {2023},
  month = {Mar},
  publisher = {American Physical Society},
  doi = {10.1103/PhysRevLett.130.121601},
  url = {https://link.aps.org/doi/10.1103/PhysRevLett.130.121601}
}

@article{I12,
author = {García Etxebarria, Iñaki},
title = {Branes and Non-Invertible Symmetries},
journal = {Fortschritte der Physik},
volume = {70},
number = {11},
pages = {2200154},
doi = {https://doi.org/10.1002/prop.202200154},
url = {https://onlinelibrary.wiley.com/doi/abs/10.1002/prop.202200154},
eprint = {https://onlinelibrary.wiley.com/doi/pdf/10.1002/prop.202200154},
year = {2022}
}

@Article{I13,
author={Niro, Pierluigi
and Roumpedakis, Konstantinos
and Sela, Orr},
title={Exploring non-invertible symmetries in free theories},
journal={Journal of High Energy Physics},
year={2023},
month={Mar},
day={01},
volume={2023},
number={3},
pages={5},
issn={1029-8479},
doi={10.1007/JHEP03(2023)005},
url={https://doi.org/10.1007/JHEP03(2023)005}
}

@article{I14,
    author = "Kaidi, Justin and Ohmori, Kantaro and Zheng, Yunqin",
    title = "{Symmetry TFTs for Non-invertible Defects}",
    eprint = "2209.11062",
    archivePrefix = "arXiv",
    primaryClass = "hep-th",
    doi = "10.1007/s00220-023-04859-7",
    journal = "Commun. Math. Phys.",
    volume = "404",
    number = "2",
    pages = "1021--1124",
    year = "2023"
}

@article{Thorngren2015,
    author = "Thorngren, Ryan and von Keyserlingk, Curt",
    title = "{Higher SPT's and a generalization of anomaly in-flow}",
    eprint = "1511.02929",
    archivePrefix = "arXiv",
    primaryClass = "cond-mat.str-el",
    month = "11",
    year = "2015"
}

@article{C1,
    author = "Kong, Liang and Wen, Xiao-Gang",
    title = "{Braided fusion categories, gravitational anomalies, and the mathematical framework for topological orders in any dimensions}",
    eprint = "1405.5858",
    archivePrefix = "arXiv",
    primaryClass = "cond-mat.str-el",
    month = "5",
    year = "2014"
}

@article{C2,
  title = {Cheshire charge in (3+1)-dimensional topological phases},
  author = {Else, Dominic V. and Nayak, Chetan},
  journal = {Phys. Rev. B},
  volume = {96},
  issue = {4},
  pages = {045136},
  numpages = {17},
  year = {2017},
  month = {Jul},
  publisher = {American Physical Society},
  doi = {10.1103/PhysRevB.96.045136},
  url = {https://link.aps.org/doi/10.1103/PhysRevB.96.045136}
}

@article{C3,
    author = "Gaiotto, Davide and Johnson-Freyd, Theo",
    title = "{Condensations in higher categories}",
    eprint = "1905.09566",
    archivePrefix = "arXiv",
    primaryClass = "math.CT",
    month = "5",
    year = "2019"
}

@article{C4,
  title = {Algebraic higher symmetry and categorical symmetry: A holographic and entanglement view of symmetry},
  author = {Kong, Liang and Lan, Tian and Wen, Xiao-Gang and Zhang, Zhi-Hao and Zheng, Hao},
  journal = {Phys. Rev. Res.},
  volume = {2},
  issue = {4},
  pages = {043086},
  numpages = {53},
  year = {2020},
  month = {Oct},
  publisher = {American Physical Society},
  doi = {10.1103/PhysRevResearch.2.043086},
  url = {https://link.aps.org/doi/10.1103/PhysRevResearch.2.043086}
}

@inbook{C5,
    author = "Johnson-Freyd, Theo",
    title = "{(3+1)D topological orders with only a Z2-charged particle}",
    eprint = "2011.11165",
    archivePrefix = "arXiv",
    primaryClass = "math.QA",
    doi = "10.1090/conm/813/16288",
    year = "2025"
}

@Article{D1,
	title={{Higher central charges and topological boundaries in 2+1-dimensional TQFTs}},
	author={Justin Kaidi and Zohar Komargodski and Kantaro Ohmori and Sahand Seifnashri and Shu-Heng Shao},
	journal={SciPost Phys.},
	volume={13},
	pages={067},
	year={2022},
	publisher={SciPost},
	doi={10.21468/SciPostPhys.13.3.067},
	url={https://scipost.org/10.21468/SciPostPhys.13.3.067},
}

@article{D2,
  title = {Noninvertible duality defects in $3+1$ dimensions},
  author = {Choi, Yichul and C\'ordova, Clay and Hsin, Po-Shen and Lam, Ho Tat and Shao, Shu-Heng},
  journal = {Phys. Rev. D},
  volume = {105},
  issue = {12},
  pages = {125016},
  numpages = {19},
  year = {2022},
  month = {Jun},
  publisher = {American Physical Society},
  doi = {10.1103/PhysRevD.105.125016},
  url = {https://link.aps.org/doi/10.1103/PhysRevD.105.125016}
}

@Article{D3,
author={Choi, Yichul
and C{\'o}rdova, Clay
and Hsin, Po-Shen
and Lam, Ho Tat
and Shao, Shu-Heng},
title={Non-invertible Condensation, Duality, and Triality Defects in 3+1 Dimensions},
journal={Communications in Mathematical Physics},
year={2023},
month={Aug},
day={01},
volume={402},
number={1},
pages={489-542},
issn={1432-0916},
doi={10.1007/s00220-023-04727-4},
url={https://doi.org/10.1007/s00220-023-04727-4}
}

@article{LUO2024,
title = {Lecture notes on generalized symmetries and applications},
journal = {Physics Reports},
volume = {1065},
pages = {1-43},
year = {2024},
note = {Lecture notes on generalized symmetries and applications},
issn = {0370-1573},
doi = {https://doi.org/10.1016/j.physrep.2024.02.002},
url = {https://www.sciencedirect.com/science/article/pii/S0370157324000528},
author = {Ran Luo and Qing-Rui Wang and Yi-Nan Wang}
}

@article{KramersWannier,
  title = {Statistics of the Two-Dimensional Ferromagnet. Part I},
  author = {Kramers, H. A. and Wannier, G. H.},
  journal = {Phys. Rev.},
  volume = {60},
  issue = {3},
  pages = {252--262},
  numpages = {0},
  year = {1941},
  month = {Aug},
  publisher = {American Physical Society},
  doi = {10.1103/PhysRev.60.252},
  url = {https://link.aps.org/doi/10.1103/PhysRev.60.252}
}

@book{Shankar,
    author = "Shankar, Ramamurti",
    title = "{Quantum Field Theory and Condensed Matter}",
    isbn = "978-1-108-36502-4, 978-0-521-59210-9",
    publisher = "Cambridge University Press",
    month = "8",
    year = "2017"
}

@article{tHooft:1979rat,
    author = "'t Hooft, Gerard",
    editor = "'t Hooft, Gerard and Itzykson, C. and Jaffe, A. and Lehmann, H. and Mitter, P. K. and Singer, I. M. and Stora, R.",
    title = "{Naturalness, chiral symmetry, and spontaneous chiral symmetry breaking}",
    reportNumber = "PRINT-80-0083 (UTRECHT)",
    doi = "10.1007/978-1-4684-7571-5_9",
    journal = "NATO Sci. Ser. B",
    volume = "59",
    pages = "135--157",
    year = "1980"
}

@misc{drawio,
author = {JGraph},
month = {10},
title = {diagrams.net, draw.io},
url = {https://www.diagrams.net/},
year = {2021}
}

@article{Cacciapaglia2026,
    author = "Cacciapaglia, Giacomo and Isnard, Wanda and Pasechnik, Roman and Preda, Anca",
    title = "{One loop renormalization of 5D gauge-Yukawa theories}",
    eprint = "2601.00453",
    archivePrefix = "arXiv",
    primaryClass = "hep-th",
    month = "1",
    year = "2026"
}

@article{Choi2024,
  title = {Quantization of Axion-Gauge Couplings and Noninvertible Higher Symmetries},
  author = {Choi, Yichul and Forslund, Matthew and Lam, Ho Tat and Shao, Shu-Heng},
  journal = {Physical Review Letters},
  volume = {132},
  number = {12},
  pages = {121601},
  year = {2024},
  doi = {10.1103/PhysRevLett.132.121601},
  eprint = {2309.03937},
  archivePrefix = {arXiv},
  primaryClass = {hep-ph},
  url = {https://link.aps.org/doi/10.1103/PhysRevLett.132.121601}
}

@article{Cordova2024-III,
  title = {Neutrino Masses from Generalized Symmetry Breaking},
  author = {C{\'o}rdova, Clay and Hong, Sungwoo and Koren, Seth and Ohmori, Kantaro},
  journal = {Physical Review X},
  volume = {14},
  number = {3},
  pages = {031033},
  year = {2024},
  doi = {10.1103/PhysRevX.14.031033},
  eprint = {2211.07639},
  archivePrefix = {arXiv},
  primaryClass = {hep-ph},
  url = {https://link.aps.org/doi/10.1103/PhysRevX.14.031033}
}

@article{Putrov2024,
  title = {Categorical Symmetry of the Standard Model from Gravitational Anomaly},
  author = {Putrov, Pavel and Wang, Juven},
  journal = {Physical Review D},
  volume = {110},
  number = {12},
  pages = {125028},
  year = {2024},
  doi = {10.1103/PhysRevD.110.125028},
  eprint = {2302.14862},
  archivePrefix = {arXiv},
  primaryClass = {hep-th},
  url = {https://link.aps.org/doi/10.1103/PhysRevD.110.125028}
}

@Article{Koren2025,
  title={{Fractionally charged particles at the energy frontier: The SM gauge group and one-form global symmetry}},
  author={Seth Koren and Adam Martin},
  journal={SciPost Phys.},
  volume={18},
  pages={004},
  year={2025},
  publisher={SciPost},
  doi={10.21468/SciPostPhys.18.1.004},
  url={https://scipost.org/10.21468/SciPostPhys.18.1.004},
  eprint={2406.17850},
  archivePrefix={arXiv},
  primaryClass={hep-ph}
}

@article{Cordova2025,
  title = {Noninvertible Peccei-Quinn Symmetry and the Massless Quark Solution to the Strong CP Problem},
  author = {C{\'o}rdova, Clay and Hong, Sungwoo and Koren, Seth},
  journal = {Physical Review X},
  volume = {15},
  number = {3},
  pages = {031011},
  year = {2025},
  doi = {10.1103/PhysRevX.15.031011},
  eprint = {2402.12453},
  archivePrefix = {arXiv},
  primaryClass = {hep-ph},
  url = {https://link.aps.org/doi/10.1103/PhysRevX.15.031011}
}

@article{Davighi2025,
  author = {Davighi, Joe},
  title = {Generalized symmetries in particle physics},
  archivePrefix = {arXiv},
  eprint = {2504.05960},
  journal = {PoS},
  volume = {DISCRETE2024},
  pages = {076},
  year = {2025},
  doi = {10.22323/1.481.0076},
  url = {https://cds.cern.ch/record/2932903}
}

@article{Dienes1998,
title = {Extra spacetime dimensions and unification},
journal = {Physics Letters B},
volume = {436},
number = {1},
pages = {55-65},
year = {1998},
issn = {0370-2693},
doi = {https://doi.org/10.1016/S0370-2693(98)00977-0},
url = {https://www.sciencedirect.com/science/article/pii/S0370269398009770},
author = {Keith R. Dienes and Emilian Dudas and Tony Gherghetta}
}

@article{Dienes1999,
title = {Grand unification at intermediate mass scales through extra dimensions},
journal = {Nuclear Physics B},
volume = {537},
number = {1},
pages = {47-108},
year = {1999},
issn = {0550-3213},
doi = {https://doi.org/10.1016/S0550-3213(98)00669-5},
url = {https://www.sciencedirect.com/science/article/pii/S0550321398006695},
author = {Keith R. Dienes and Emilian Dudas and Tony Gherghetta}
}

@article{Bhattacharyya2007,
title = {Power law scaling in universal extra dimension scenarios},
journal = {Nuclear Physics B},
volume = {760},
number = {1},
pages = {117-127},
year = {2007},
issn = {0550-3213},
doi = {https://doi.org/10.1016/j.nuclphysb.2006.10.018},
url = {https://www.sciencedirect.com/science/article/pii/S0550321306008479},
author = {Gautam Bhattacharyya and Anindya Datta and Swarup {Kumar Majee} and Amitava Raychaudhuri}
}

@Article{Whitehead1949,
author={Whitehead, J. H. C.},
title={On simply connected, 4-dimensional polyhedra},
journal={Commentarii Mathematici Helvetici},
year={1949},
month={Dec},
day={01},
volume={22},
number={1},
pages={48-92},
issn={1420-8946},
doi={10.1007/BF02568048},
url={https://doi.org/10.1007/BF02568048}
}

@article{Bucci2003,
  title = {Effective potential and vacuum stability within universal extra dimensions},
  author = {Bucci, Patrizia and Grzadkowski, Bohdan},
  journal = {Phys. Rev. D},
  volume = {68},
  issue = {12},
  pages = {124002},
  numpages = {12},
  year = {2003},
  month = {Dec},
  publisher = {American Physical Society},
  doi = {10.1103/PhysRevD.68.124002},
  url = {https://link.aps.org/doi/10.1103/PhysRevD.68.124002}
}

@article{DelZotto:2024ngj,
    author = "Del Zotto, Michele and Dell'Acqua, Matteo and Riedel G{\r{a}}rding, Elias",
    title = "{The Higher Structure of Symmetries of Axion-Maxwell Theory}",
    eprint = "2411.09685",
    archivePrefix = "arXiv",
    primaryClass = "hep-th",
    doi = "10.1007/s00220-026-05671-9",
    journal = "Commun. Math. Phys.",
    volume = "407",
    number = "8",
    pages = "170",
    year = "2026"
}

@article{Davighi:2024zjp,
    author = "Davighi, Joe and Lohitsiri, Nakarin",
    title = "{WZW terms without anomalies: Generalised symmetries in chiral Lagrangians}",
    eprint = "2407.20340",
    archivePrefix = "arXiv",
    primaryClass = "hep-th",
    reportNumber = "CERN-TH-2024-120",
    doi = "10.21468/SciPostPhys.17.6.168",
    journal = "SciPost Phys.",
    volume = "17",
    number = "6",
    pages = "168",
    year = "2024"
}

@article{Adachi:2021rjw,
    author = "Adachi, Yuki and Lim, C. S. and Maru, Nobuhito",
    title = "{The strong CP problem and higher-dimensional gauge theories}",
    eprint = "2108.07367",
    archivePrefix = "arXiv",
    primaryClass = "hep-th",
    reportNumber = "OCU-PHYS-543, NITEP-111",
    doi = "10.1093/ptep/ptac070",
    journal = "PTEP",
    volume = "2022",
    number = "5",
    pages = "053B06",
    year = "2022"
}

\end{document}